\documentclass[sigconf, nonacm]{acmart}

\AtBeginDocument{%
  }

\usepackage{amsmath} 
\usepackage{balance} 
\usepackage{booktabs} 
\usepackage{caption}
\usepackage{comment}
\usepackage{enumitem}
\setlist[itemize]{leftmargin=2.2em}

\usepackage{fancyvrb} 
\usepackage[symbol]{footmisc}
\usepackage{graphicx} 
\usepackage{makecell} 
\usepackage{multirow} 
\usepackage{soul}
\usepackage{subfigure} 
\usepackage{xcolor}
\usepackage{xspace} 
\usepackage{bm}
\usepackage{utfsym}
\usepackage{cancel}
\usepackage[linesnumbered,ruled,vlined]{algorithm2e}
\usepackage{pifont}
\usepackage{tikz}
\usepackage{marginnote}
\usepackage{mparhack}

\theoremstyle{definition}

\theoremstyle{plain}

\theoremstyle{remark}

\renewcommand{\paragraph}[1]{\vspace{0.1em}
\noindent\textbf{#1.}}
\renewcommand{\subparagraph}[1]{\vspace{0.1em}
\noindent\textit{\underline{#1.}}}

\newcommand\dbname{\ensuremath{\textsf{NeurStore}}\xspace}

\usepackage{multicol}

\usepackage{wrapfig}
\usepackage{listings}
\usepackage[most]{tcolorbox}

\newtcolorbox[auto counter]{mybox}[1][]{%
breakable,
enhanced,
sharp corners,
colback=white,
fonttitle=\bfseries,
enlarge bottom at break by=5mm,
enlarge top at break by=5mm,
overlay first={%
    \draw[black, line width=0.5mm](frame.south west)--(frame.south east);
    \node[anchor=north east] at (frame.south east) {continued on next page};
    },
overlay middle={%
    \draw[black, line width=0.5mm](frame.south west)--(frame.south east);
    \draw[black, line width=0.5mm](frame.north west)--(frame.north east);
    \node[anchor=north east] at (frame.south east) {continued on next page};
    \node[anchor=south west] at (frame.north west) {continued from next page};
    },
overlay last={%
    \draw[black, line width=0.5mm](frame.north west)--(frame.north east);
    \node[anchor=south west] at (frame.north west) {continued from next page};},
#1
}

  {\list{}{\leftmargin=0.15in\rightmargin=0.15in}\item[]}%
  {\endlist}

\newtcolorbox{myquote}[1][]{
    colback=black!10,
    colframe=black!10,
    notitle,
    sharp corners,
    enhanced,
    breakable,
    left=2pt,
    right=2pt,
    top=2pt,
    bottom=2pt,
    ignore nobreak,
}

\definecolor{R1color}{HTML}{e66101} 
\definecolor{R2color}{HTML}{7b3294} 
\definecolor{R4color}{HTML}{018571} 
\newcommand{\revone}[1]{\textcolor{R1color}{#1}}

\newcommand{\revfour}[1]{\textcolor{R4color}{#1}}

\newtcolorbox{myquoteMeta}[1][]{
    colback=black!3,
    colframe=black!3,
    notitle,
    sharp corners,
    borderline west={1.5pt}{0pt}{blue},
    enhanced,
    breakable,
    left=2pt,
    right=2pt,
    top=2pt,
    bottom=2pt,
    ignore nobreak,
}

\newtcolorbox{myquoteR1}[1][]{
    colback=black!3,
    colframe=black!3,
    notitle,
    sharp corners,
    borderline west={1.5pt}{0pt}{R1color},
    enhanced,
    breakable,
    left=2pt,
    right=2pt,
    top=2pt,
    bottom=2pt,
    ignore nobreak,
}

\newtcolorbox{myquoteR2}[1][]{
    colback=black!3,
    colframe=black!3,
    notitle,
    sharp corners,
    borderline west={1.5pt}{0pt}{R2color},
    enhanced,
    breakable,
    left=2pt,
    right=2pt,
    top=2pt,
    bottom=2pt,
    ignore nobreak,
    #1
}

\newtcolorbox{myquoteR4}[1][]{
    colback=black!3,
    colframe=black!3,
    notitle,
    sharp corners,
    borderline west={1.5pt}{0pt}{R4color},
    enhanced,
    breakable,
    left=2pt,
    right=2pt,
    top=2pt,
    bottom=2pt,
    ignore nobreak,
}

\newif\ifextended\extendedfalse

\newcommand{\maintext}[1]{\ifextended\relax\else#1\fi} 

\newcommand{\extended}[1]{\ifextended#1\else\relax\fi}

  {\endminipage\par\medskip}

\renewcommand{\thefootnote}{}
\thanks{$^*$ Corresponding author.}
\renewcommand{\thefootnote}{\arabic{footnote}}

\begin{document}



\maintext{\title{{\dbname}: Efficient In-database Deep Learning Model Management System}}

\extended{\title{{\dbname}: Efficient In-database Deep Learning Model Management System (Extended Version)}}


\author{
Siqi Xiang$^1$, Sheng Wang$^2$, Xiaokui Xiao$^1$, Cong Yue$^{1*}$\thefootnote{}, Zhanhao Zhao$^1$, Beng Chin Ooi$^3$
}

\affiliation{
\fontsize{10}{10}\textit{$^1$ National University of Singapore}
\qquad	\fontsize{10}{10}\textit{$^2$ Alibaba Group}
\qquad	\fontsize{10}{10}\textit{$^3$ Zhejiang University} \\
\fontsize{9}{9}{\{siqxiang, xiaoxk, yuecong, zhzhao\}@comp.nus.edu.sg} \qquad
\fontsize{9}{9}{sh.wang@alibaba-inc.com} \qquad 
\fontsize{9}{9}{ooibc@zju.edu.cn} \qquad 
\country{}
}

\renewcommand{\shortauthors}{Siqi Xiang, Sheng Wang, Xiaokui Xiao, Cong Yue, Zhanhao Zhao, Beng Chin Ooi}

\sloppy






\begin{abstract}
With the prevalence of in-database AI-powered analytics, there is an increasing demand for database systems to efficiently manage the ever-expanding number and size of deep learning models.
However, existing database systems typically store entire models as monolithic files or apply compression techniques that overlook the structural characteristics of deep learning models, resulting in suboptimal model storage overhead.
This paper presents \dbname, a novel in-database model management system that enables efficient storage and utilization of deep learning models.
First, \dbname employs a tensor-based model storage engine to enable fine-grained model storage within databases.
In particular, we enhance the hierarchical navigable small world (HNSW) graph to index tensors, and only store additional deltas for tensors within a predefined similarity threshold to ensure tensor-level deduplication.
Second, we propose a delta quantization algorithm that effectively compresses delta tensors, thus achieving a superior compression ratio with controllable model accuracy loss.
Finally, we devise a compression-aware model loading mechanism, which improves model utilization performance by enabling direct computation on compressed tensors.
Experimental evaluations demonstrate that \dbname achieves superior compression ratios and competitive model loading throughput compared to state-of-the-art approaches.

\end{abstract}



\maketitle

\section{Introduction} \label{sec:introduction}


Modern database management systems (DBMSs) are increasingly integrating artificial intelligence (AI) to support advanced data analytics~\cite{postgresml,VerticaML_SIGMOD2020,EVA_SIGMOD22,RAVEN_SIGMOD22,NeurDB,DBLP:journals/pvldb/XingCCLOP24}.
Such in-database AI-powered analytics enable users to issue complex data analytics tasks through specialized SQL interfaces~\cite{DBLP:conf/sigmod/ZhangPXXY24,DBLP:journals/pacmmod/ZhangPXXY25,NeurDB}.
DBMSs then automatically retrieve the relevant stored data and perform AI inference to provide deeper insights that traditional statistical operations (e.g., averages and sums) often fail to capture.
As a result, the entire AI analytic workflow occurs within the database, which eliminates the need to move large amounts of data outside DBMSs, and thus facilitates efficient and secure analytics~\cite{VerticaML_SIGMOD2020,DBLP:conf/sigmod/XuQYJRGKLL0Y022}.


Sectors such as finance~\cite{DBLP:journals/pvldb/HellersteinRSWFGNWFLK12,DBLP:conf/sigmod/FengKRR12} and e-commerce~\cite{RAVEN_SIGMOD22,NeurDB} are rapidly adopting in-database AI-powered analytics in their critical business workflows, and with the advancements of AI, deep learning (DL) models have become prevalent.
Consider purchase recommendations in e-commerce as an illustrative example, where items are recommended based on personal user profiles, 
including habits, occupations, and lifestyles.
As user profiles typically contain sensitive information, such as salary and browsing history, in-database analytics is particularly suitable for handling such recommendation tasks.
To enable precise and personalized recommendations, e-commerce vendors commonly deploy specialized DL models tailored to different users, regions, or customer segments.
These specialized models are often derived from fundamental pre-trained DL models~\cite{DBLP:journals/pvldb/ZengXCCOPW24,ooi2015singa,luo2021mlcask,gao2023enabling}, which may consist of dozens or even hundreds of layers, each potentially requiring gigabytes of storage~\cite{comp_Elves_VLDB2024,ModelDB_SIGMOD2016,ModelHub_ICDE2017}.
Further, as user profiles continuously evolve over time, these specialized models must be regularly updated or fine-tuned, leading to a steadily increasing number of models.
As a result, efficient in-database DL model management (i.e., storing and loading DL models directly within DBMSs) has become a foundational capability for in-database AI-powered analytics.

Existing in-database model management approaches generally treat each model as an isolated unit and store full-fledged models independently.
For example, DBMSs such as PostgresML~\cite{postgresml}, Oracle~\cite{oracle}, and Azure~\cite{azure} serialize each model into a BLOB and store it directly in a dedicated model table~\cite{postgresml,oracle,azure}.
Alternatively, systems such as ModelDB~\cite{ModelDB_SIGMOD2016}, RAVEN~\cite{RAVEN_SIGMOD22}, and Vertica-ML~\cite{VerticaML_SIGMOD2020} store file paths in the table, while placing the actual models as external files.
Although straightforward, these methods suffer from substantial storage overhead, as storing hundreds of models can require terabytes of space due to redundant parameters and deep architectures. This overhead grows rapidly with the scale and complexity of DL model deployments.
To mitigate storage costs, users can manually compress models before storing them into DBMSs using general data compression algorithms~\cite{zstandard,zfp,zlib}, floating-point compression schemes~\cite{zfp}, or specialized model compression methods~\cite{comp_Elves_VLDB2024}.
However, such optimization only partially addresses the storage issue, as they still handle each model independently, i.e., the overall storage cost remains proportional to the total number and size of models~\cite{comp_Elves_VLDB2024}.
This persistent linear growth in storage overhead presents a critical bottleneck for scalable DL model management. Addressing this challenge requires new strategies that exploit structural similarities across models to reduce redundancy.

Storing a DL model involves two core components: 1) the model architecture, typically represented as a computational graph that defines the connectivity and operations of layers; and 2) a set of layers, where each layer comprises one or more high-dimensional floating-point tensors.
As model fine-tuning is relatively common, many DL models share similar architectures and contain identical or highly similar tensors, particularly when fine-tuning is limited to a subset of layers~\cite{DBLP:conf/iclr/HuSWALWWC22,ModelHub_ICDE2017,comp_Elves_VLDB2024}.
This observation naturally motivates us to explore similarities and relationships across models, thus eliminating redundancy and improving overall storage efficiency.


Given that the model's learnable parameters (e.g., weights and biases) in tensors constitute the majority of a model's storage cost, 
we aim to identify redundant tensors and store only incremental differences between similar tensors.
However, achieving this tensor-level deduplication requires addressing three key challenges:
First, similarities between tensors are implicit.
For instance, two models without explicit lineage may still contain similar tensors.
Consequently, effectively identifying similar layers across a large collection of models is inherently challenging.
Second, tensors typically consist of high-entropy floating-point parameters.
Even if two tensors exhibit similar structures and parameters, directly calculating their parameter-wise differences can result in a delta tensor of identical dimensionality, yielding little to no storage savings.
Thus, generating compact delta tensors that meaningfully reduce storage consumption is not straightforward.
Third, since models are stored as fine-grained delta tensors, retrieving a model involves reconstructing the complete model based on these tensors.
Due to the complexity and depth of DL models, this reconstruction process can be costly, and therefore, efficiently retrieving models is non-trivial.

In this paper, we present \dbname, an efficient in-database model management system designed to reduce DL model storage costs while streamlining model utilization.
We first propose a tensor-based storage engine that departs from traditional per-model storage by identifying and storing shared tensor components across models, significantly improving space efficiency through structured deduplication.
At its core is a high-performance tensor index built upon the Hierarchical Navigable Small World (HNSW) graph structure.
Specifically, we categorize tensors into two types: \textsl{base tensors}, which store original parameters, and \textsl{delta tensors}, which maintain differences relative to a corresponding base tensor.
Base tensors are stored as nodes in the HNSW-based tensor index, while delta tensors are placed separately in dedicated tensor pages.
When saving a new model, we deconstruct it into individual tensors, and for each tensor, we search the tensor index to determine whether a similar base tensor already exists within a predefined similarity threshold.
If such a base tensor is found, we compute and store the corresponding delta tensor; otherwise, unmatched tensors are stored as new base tensors.
To integrate this design seamlessly into DBMSs, we build the tensor-based storage engine on top of modern database architecture, with enhancements specifically designed for efficient DL model management.
In particular, we introduce a tailored index cache that efficiently buffers portions of the HNSW-based tensor index and extend the native page layout to support large tensors without disrupting the existing page-based storage mechanism.

We then introduce a delta quantization algorithm that compresses delta tensors to achieve reduced storage overhead.
Unlike traditional quantization that operates on complete models, we quantize delta tensors, whose parameter ranges are typically much narrower than those of the original tensors.
This mitigates the accuracy loss commonly associated with traditional model quantization.
Moreover, our algorithm is adaptive, dynamically selecting the bit width for each delta tensor based on its parameter distribution, enabling fine-grained control over the trade-off between storage efficiency and model accuracy.

Lastly, we design a compression-aware model loading mechanism that enables direct computation on compressed tensors, eliminating the need to fully reconstruct models before use.
Unlike traditional pipelines that first decompress models into memory, our approach integrates the reconstruction directly into the computation graph and pipelines tensor loading with computation. 
This tight integration reduces inference latency and memory overhead, improving model loading performance in in-database settings.

In summary, we make the following contributions:

\begin{itemize}[leftmargin=*]

\item We present \dbname, a novel in-database DL model management system that enables efficient tensor-level storage and loading.

\item We introduce a structured tensor-based storage engine that can be seamlessly integrated into modern DBMSs.

\item We develop an adaptive delta quantization algorithm that minimizes storage by dynamically adjusting the bit width for each delta tensor.

\item We design a compression-aware model loading mechanism that supports direct computation over compressed tensors, reducing the overall in-database AI-powered analytics latency.

\item We implement \dbname as a pluggable PostgreSQL extension and evaluate its performance against state-of-the-art systems.
Experimental results demonstrate substantial gains in end-to-end AI-powered analytics performance, storage efficiency, and model saving and loading throughput.
%
We further integrate \dbname into two analytical databases~\cite{DBLP:conf/sigmod/RaasveldtM19,DBLP:journals/pvldb/SchulzeSYDM24}, and the performance evaluation confirms its general extensibility.


\end{itemize}


\dbname is a component of NeurDB~\cite{NeurDB,DBLP:journals/corr/abs-2408-03013}.  It has been integrated into NeurDB as its native model storage.

The remainder of the paper is structured as follows. 
Section~\ref{sec:background} provides the relevant background and presents the problem statement.
Section~\ref{sec:overview} overviews the system architecture of \dbname.
Section~\ref{sec:design} details the design of \dbname.
Section~\ref{sec:implementation} describes the system implementation, and Section~\ref{sec:evaluation} presents the experimental results.
Section~\ref{sec:related} reviews the related works, and finally, Section~\ref{sec:conclusion} concludes the paper.

\section{Background}
\label{sec:background}

In this section, we provide the relevant background and formally define the problem that \dbname aims to address.

\subsection{In-database AI-powered Analytics}
\label{subsec:bg_indb_ai}

\begin{figure}
\centering
\includegraphics[width=0.96\linewidth]{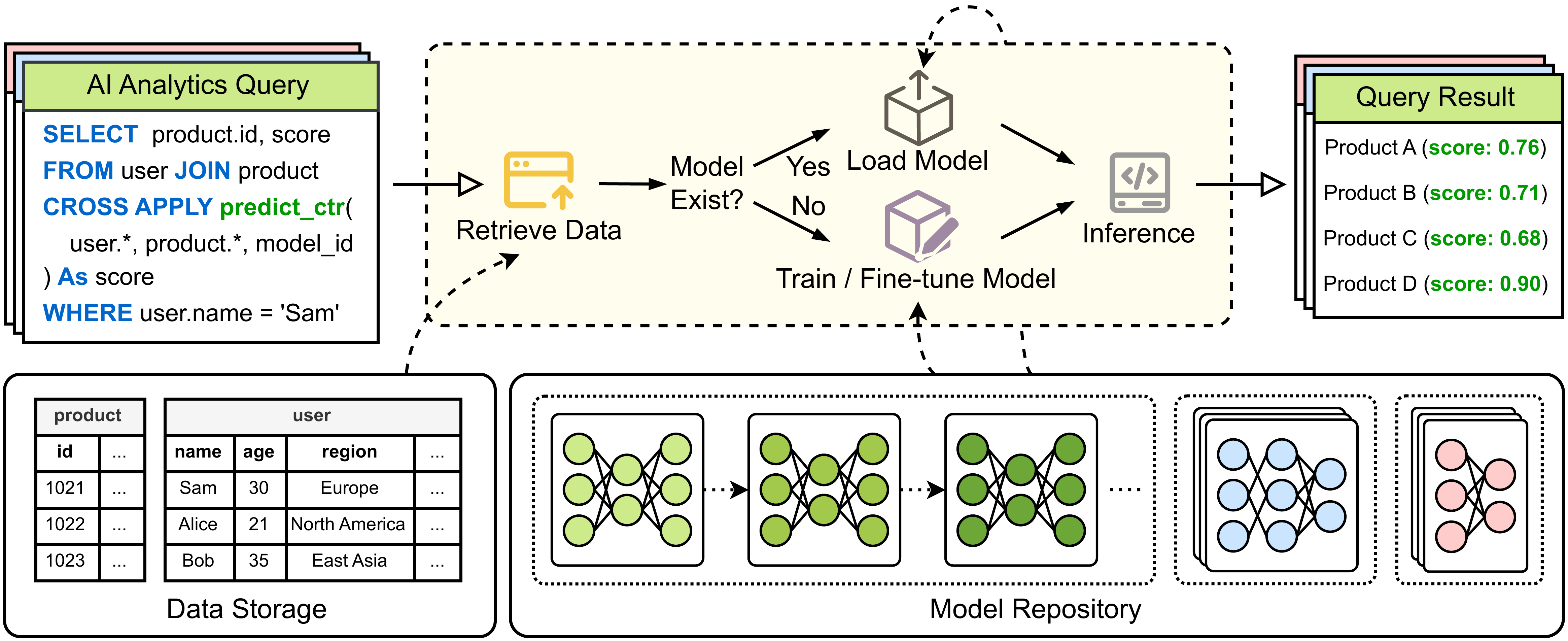}
\vspace{-2mm}
\caption{
In-database AI-powered Analytics Workflow {\rm -- It consists of three main steps: (1) data retrieval, (2) model loading, and (3) inference.}
}
\label{fig:in_db_ai_workflow}
\end{figure}

In-database AI-powered analytics enables DBMSs to handle complex data analytics tasks via specialized SQL syntax~\cite{NeurDB,oracle} or user-defined functions (UDFs)~\cite{postgresml,EVA_SIGMOD22}.
For example, let us consider the click-through rate (CTR) prediction task, as shown in Figure~\ref{fig:in_db_ai_workflow}.
A data analyst submits a query to estimate CTR scores, i.e., the probability of a user clicking on the product.
Upon receiving the query, the DBMS retrieves relevant data (e.g., product ID, user name, etc.) from the user and product tables, loads the DL model specified by the model ID, and performs model inference to predict CTR scores.



Since different analytics tasks often require different DL models, a model repository is typically maintained, where models are stored as files or binary large objects (BLOBs). 
Each model may have dozens or hundreds of layers and consume gigabytes of storage, and therefore, repeatedly loading these large models from disk for different tasks can introduce substantial latency.
Moreover, with ongoing data updates, models need to be frequently retrained or fine-tuned to maintain accuracy.
For instance, a CTR prediction model may generate multiple updated versions over time to reflect new user and product data, which significantly increases storage overhead.
These factors limit the scalability of in-database AI-powered analytics and degrade the response time of analytics queries.
To address these challenges, we design \dbname, an in-database model management system that reduces model storage overhead and improves model loading efficiency.

\subsection{Storage Optimizations for DL Models}

Existing storage optimization techniques for DL models can be broadly categorized into general-purpose data compression algorithms and model-specific compression techniques.
General compression algorithms, such as ZSTD~\cite{zstandard}, can be directly applied to serialized DL model files, and are effective at removing exact duplicate patterns in the data.
However, DL models rarely contain such duplicates~\cite{ModelHub_ICDE2017,comp_Elves_VLDB2024}, especially across layers with diverse weights.
Similarly, floating-point compression schemes such as ZFP~\cite{zfp} perform well on structured, spatially local data, but not on DL models whose weights are typically high-dimensional, continuous, broadly distributed, and lack spatial regularity. These characteristics limit the effectiveness of generic compressors.

Model-specific compression techniques address the aforementioned shortcomings.
In particular, ELF~\cite{comp_Elves_VLDB2024} eliminates the exponent bits of floating-point values within the range (–1, 1) by remapping them to the interval [1, 2), to improve compressibility.
Nonetheless, ELF operates on each model in isolation, missing opportunities to exploit shared structure across models.

Rather than optimizing each model independently, we aim to reduce storage costs by leveraging inter-model similarities and shared components. 
\dbname builds on this idea by enabling tensor-level deduplication, identifying and reusing redundant model components across a collection.

\subsection{Hierarchical Navigable Small World}
\label{sec:bg_hnsw}

Approximate Nearest Neighbor search (ANN) is a technique for efficiently identifying data points in high-dimensional spaces that are approximately closest to a given query point~\cite{similarity_survey_A,similarity_faiss}.
Existing ANN algorithms can be categorized into hashing-based~\cite{similarity_LSH_A,similarity_LSH_B}, tree-based~\cite{similarity_tree}, quantization-based~\cite{similarity_quant_A,similarity_quant_B}, and graph-based approaches~\cite{similarity_HNSW,similarity_NSG}. 
Among these, the Hierarchical Navigable Small World (HNSW) graph~\cite{similarity_HNSW}, a state-of-the-art graph-based method, is widely adopted due to its ability to achieve high recall with low query latency.
HNSW organizes data points into a multi-layer graph structure.
Each layer forms a navigable small-world graph, where nodes are connected to their approximate nearest neighbors.
Higher layers provide coarse-grained shortcuts, while the lower layers enable fine-grained local search.
Given a query, the search algorithm starts from a high-level node and performs greedy search layer by layer, descending through the hierarchy until it reaches the bottom layer, where it refines the search to identify the approximate nearest neighbor.
Formally, let $G = (V, E)$ denote the HNSW graph, where $V$ is the set of data points represented as vertices, and $E$ consists of edges between points, weighted by their pairwise distances.
Given a query point $q$ and an entry point $v_0$, HNSW iteratively traverses the graph to find the neighbor $v_{t+1}$ of the current vertex $v_t$ that is closest to $q$.
This process continues until no closer neighbor is found, at which point the current node is returned as the approximate nearest neighbor.

In \dbname, we leverage HNSW to index base tensors efficiently.
This allows us to identify previously stored base tensors that are most similar to a given input tensor.

\subsection{Post-training Model Quantization}


Quantization is designed to reduce the computational and memory costs of DL models~\cite{comp_quant_A,comp_quant_B, comp_quant_C}.
It maps a model's weights and activations to lower-precision formats (e.g., from Float32 to Int8).
In general, quantization techniques can be classified into quantization-aware training (QAT)~\cite{comp_quant_QAT_A,comp_quant_QAT_B} and post-training quantization (PTQ)~\cite{comp_quant_PTQ_A,comp_quant_PTQ_B,DBLP:GOBO}. 
QAT integrates quantization operations during model training, while PTQ is performed after model training.
In PTQ, a Float32 tensor $\theta=\{x_1,x_2,...,x_n\}$, where $x_i\in[x_{min}, x_{max}]$ is quantized to a tensor of $b$-bit integers ${q_1, q_2, \dots, q_n}$ using a scale factor $s = \frac{x_{\max} - x_{\min}}{2^b - 1}$ and a zero-point offset $z$.
Each value is quantized as $q_i = round\left(\frac{x_i-x_{min}}{s}\right)+z$.
While PTQ is simple and efficient, it inevitably introduces precision loss, especially when the dynamic range of $X$ is wide or the bit width $b$ is small. 


Unlike traditional use cases of PTQ, we leverage PTQ in the context of model management.
We observe that delta tensors, i.e., the differences between similar tensors across models, typically exhibit smaller value ranges than the original tensors.
As a result, applying PTQ to delta tensors can help mitigate the accuracy degradation typically associated with quantizing full tensors.
To further reduce precision loss, \dbname dynamically adjusts the quantization bit width $b$ for each delta tensor, based on its value distribution and a user-defined accuracy tolerance.
This enables fine-grained compression while preserving model performance.



\subsection{Problem Definition} \label{subsec:motivation}
Modern DBMSs increasingly support AI-powered analytics by incorporating DL models.
In this context, in-database model management refers to the capability of DBMSs to store and utilize DL models efficiently.
A well-designed system must achieve the following three objectives:
(1) Storage consumption: Minimize the total storage required to maintain a collection of models, especially as the number and size of models grow.
(2) Model accuracy: Preserve the predictive performance of models, particularly when storage-saving techniques (e.g., quantization or compression) are applied.
(3) Query efficiency: Ensure high-throughput model loading and low-latency inference to support analytical workloads.

However, these objectives are often in tension with one another. 
For example, aggressive compression may reduce storage overhead, but at the cost of increased precision loss or additional decoding overhead during inference. 
Similarly, techniques that improve loading efficiency may require storing uncompressed or partially compressed models, thereby increasing storage consumption.
Given these trade-offs, it is typically infeasible to optimize all three objectives simultaneously~\cite{netsDB_VLDB2022,EVA_SIGMOD22}.
In this work, we focus on minimizing storage consumption, which becomes increasingly critical as the number and size of models grow, while maintaining acceptable model accuracy and reasonable query efficiency. 
In particular, we ensure that the accuracy degradation caused by our approach remains within a user-defined tolerance, and the model is efficient for retrieval and inference.
We now formally define the problem addressed by \dbname.

\paragraph{Problem definition}
Let $\mathcal{M} = \{M_1, M_2, \dots, M_n\}$ denote a set of $n$ DL models, where each model $M_i$ consists of a set of $L_i$ layers, i.e., $M_i = \{ \ell_{i,1}, \ell_{i,2}, \dots, \ell_{i,L_i} \}$.
Each layer $\ell_{i,j}$ contains a set of $K_{i,j}$ learnable tensors:
$\ell_{i,j} = \left\{ \theta_{i,j}^{(1)}, \theta_{i,j}^{(2)}, \dots, \theta_{i,j}^{(K_{i,j})} \right\}, \quad \theta_{i,j}^{(k)} \in \mathbb{R}^{d_{i,j}^{(k)}}$, where $d_{i,j}^{(k)}$ refers to the dimensionality of the $k$-th learnable tensor in layer $j$ of model $M_i$.
Given a precision loss tolerance $p$ (e.g., a relative or absolute error bound), the goal of \dbname is to jointly minimize the total storage cost and query latency of the DL model collection, while ensuring that the compression-induced accuracy loss remains within acceptable bounds.
Formally, let $S(\mathcal{M})$ denote the total storage cost of the model collection $\mathcal{M}$, and $T(\mathcal{M})$ denote the total query latency (e.g., model loading and inference time). The optimization objective can be defined as:
\begin{equation}
\begin{aligned}
&\min{\quad \alpha \cdot S(\mathcal{M}) + \beta \cdot T(\mathcal{M})}, \\
&\text{subject to} \quad \forall M_i \in \mathcal{M}, \; \text{Error}(M_i) \leq p.
\end{aligned}
\end{equation}


Here, $\alpha$ and $\beta$ are user-defined weights that balance the importance of storage efficiency and query performance.
$\text{Error}(M_i)$ represents the quantifiable accuracy degradation introduced by compressing model $M_i$, such as layer-wise tensor deviation or loss in downstream prediction accuracy.

\section{System Overview} \label{sec:overview}

\begin{figure}
    \centering
    \includegraphics[width=0.99\linewidth]{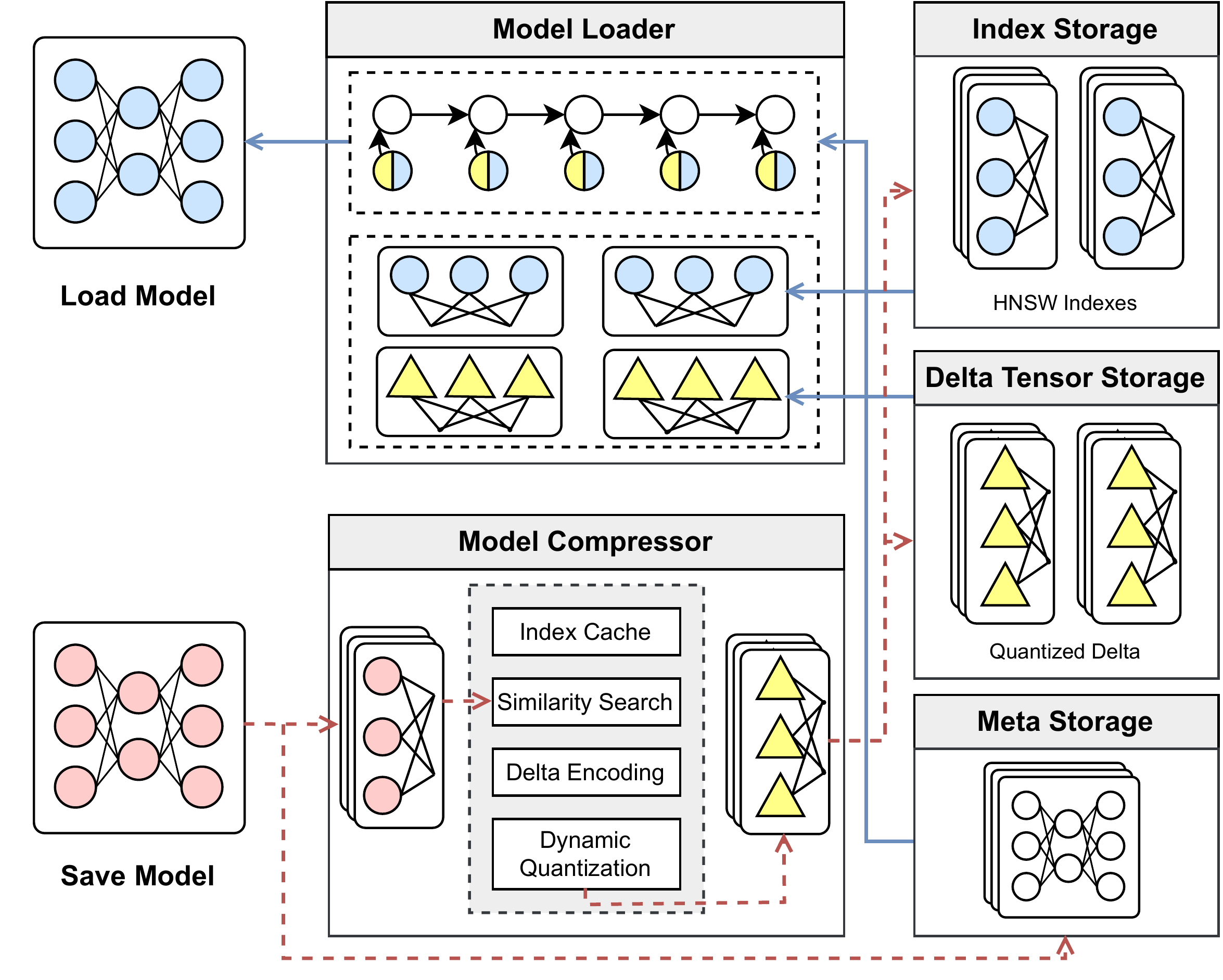}
    \vspace{-2mm}
    \caption{System Architecture of {\dbname} {\rm -- The system supports two main workflows: 
    (1) {model saving} (\textcolor[HTML]{B85450}{red}), where models are compressed before being stored, and (2) {model loading} (\textcolor[HTML]{6C8EBF}{blue}), where models are retrieved by the model loader.}}
    \label{fig:architecture}
\end{figure}

In this section, we describe the system architecture of \dbname. As shown in Figure~\ref{fig:architecture},
\dbname comprises three core components, namely the model compressor, the model loader, and the storage layer.
Upon receiving users' {\it Save model} requests, the model compressor is invoked to reduce the model size using our delta quantization algorithm.
The compressed tensor, updated ANN indexes, and model architecture are then serialized and stored in the storage layer.
During model inference, users send {\it Load model} requests to the model loader, which fetches the compressed tensors from the storage layer and performs computation without full decompression.
Next, we describe each component in detail.

\paragraph{Storage Layer}
The storage layer is responsible for managing all model-related data in \dbname, including HNSW, tensors, and model architectures.
\dbname persists HNSW on disk, which are loaded into memory at runtime and used by the model compressor for efficient similarity search.
To reduce the size of HNSW, we store the 8-bit quantized tensors in the vertices as the base tensors.
\dbname stores model tensors as quantized deltas with respect to the corresponding base tensors.
The quantization parameters, such as zero point and scale, are serialized and stored as the prefix of each quantized delta and tensor.
Additionally, the serialized model architectures are stored in the meta storage.
\dbname uses a relational table to organize model metadata, including model IDs and names, so that external users can easily manage and interact with models in the database.

\paragraph{Model Compressor} 
The model compressor is used to reduce the model size using our delta quantization algorithm (Section~\ref{subsec:compressor}) according to the following steps.
(1) The system first decouples model weights from the model architecture. 
This separation allows \dbname to manage weights at the granularity of individual tensors.
(2) A similarity search is performed for each tensor to locate the most similar base tensor in the system. 
(3) The delta between the input tensor and the base tensor is then computed.
(4) If the delta is sufficiently small to be quantized within the user-defined bit width, it is stored in the storage layer.
Otherwise, a new HNSW vertex is created using the quantized value of the input tensor, and the process repeats from Step (2).

\paragraph{Model Loader} 
The model loader is designed to efficiently retrieve the required models using the compression-aware model loading mechanism.
Specifically, \dbname loads the base tensors, delta tensors, and the computation graph into memory.
To facilitate fast retrieval, tensors are loaded without full decompression, i.e., de-quantization and reconstruction.
The shared base tensor is loaded only once, even when referenced by multiple layers or models.
To reduce the memory usage, we modify the computation graph upon retrieval so that tensors are only de-quantized and reconstructed before they are invoked in the computation.
\dbname then follows the modified computation graph to compute the results.

\paragraph{Running Example}
Let us continue the CTR prediction example shown in Figure~\ref{fig:in_db_ai_workflow}.
After receiving the analytics task, the database retrieves relevant data and initiates a {\it Load Model} request to \dbname.
In particular, \dbname employs the model loader to perform compression-aware loading, retrieving the associated tensors and computation graph for in-database CTR prediction.
Further, when a newly trained or fine-tuned model needs to be stored, the database issues a {\it Save Model} request.
In response, \dbname compresses the model using the adaptive delta quantization algorithm and persists the compressed tensors in the storage layer.

\section{Design of \dbname} \label{sec:design}

In this section, we detail the key techniques proposed for \dbname, including a tensor-based storage engine, a delta quantization algorithm, and a compression-aware model loading and inference mechanism.

\begin{figure}
    \centering
    \includegraphics[width=0.86\linewidth]{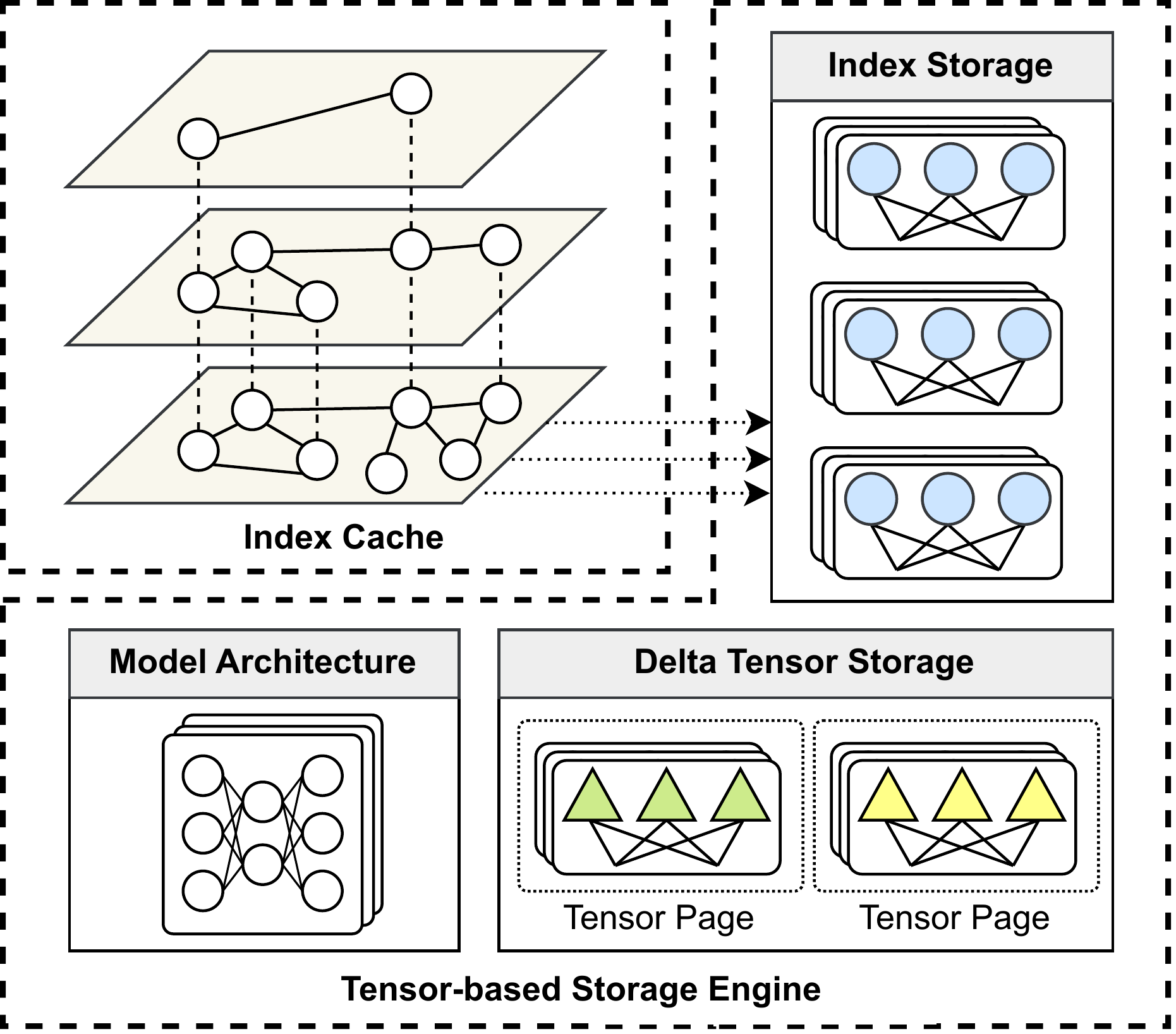}
    \vspace{-2mm}
    \caption{Tensor-based Storage Engine {\rm -- The engine consists of three components: index storage for HNSW-based indexes, delta tensor storage, and metadata storage for model architectures.}}
    \label{fig:storage_engine}
\end{figure}

\subsection{Tensor-based Storage Engine} \label{subsec:storage}

To exploit the structural similarities across DL models, \dbname organizes and compresses deep learning models at the granularity of individual tensors rather than entire models.
\extended{This design enables similarity-based delta compression across models, allowing the system to avoid redundant storage by referencing previously stored tensors. }
The overall storage layout is illustrated in Figure~\ref{fig:storage_engine}. \dbname separates model storage into two main components: index storage, which stores shared base tensors used for reference, and delta tensor storage, which stores the differences between compressed tensors and their matched references.
In addition, we serialize each model’s architecture into a dedicated metadata storage, which extends the native tablespace that stores table structures commonly used in DBMSs.


\paragraph{Index Storage}
Given that different models may contain tensors of varying shapes, \dbname maintains a collection of HNSW indexes, one per unique tensor structure (i.e., shape).
Each HNSW index organizes similarly shaped tensors into a proximity graph, where each node stores a base tensor, and edges connect similar tensors to facilitate efficient ANN search.
To reduce the index size, each base tensor is quantized to 8-bit using linear quantization prior to insertion.
Although quantization introduces some loss, \dbname preserves full-precision representation recoverability by storing a corresponding delta tensor that captures the difference between the original tensor and its quantized representation.
Our proposed delta quantization algorithm will be detailed in Section~\ref{subsec:compressor}.

\paragraph{Delta Tensor Storage}
The delta tensor storage is responsible for efficiently storing the compressed differences between base tensors and the tensors compressed relative to them.
To support the typically large size of tensor data, we introduce a new page type in the database called a tensor page. 
Unlike standard heap pages, tensor pages are allowed to exceed the traditional page size limit and are managed separately to optimize read/write performance for large blocks.
Within each tensor page, we store compressed deltas compactly.
For each delta tensor, we store the following:
1) A 4-bit scale;
2) A 4-bit zero-point;
3) A quantized weight array.
Each tensor is dynamically quantized based on its value range, and therefore, maintains its own scale and zero-point, which are used to de-quantize the delta tensor (Section~\ref{subsec:compressor}). 
The quantized weights represent the difference between a base tensor and its corresponding variant, as determined by approximate similarity search.
This design allows multiple models to share common base tensors while storing only the compressed deltas for fine-tuned variants.
To further optimize model loading performance, delta tensors are organized in the order defined by the model architecture. 
This improves spatial locality and supports efficient reconstruction during model loading and inference.

\paragraph{Index Cache}
To reduce the overhead of accessing HNSW in model compression and loading, we introduce an index cache that stores the deserialized HNSWs in memory.
When a lookup of a base tensor is invoked, the system first checks the cache; if the corresponding HNSW index exists, the system bypasses disk I/O and the de-serialization process.
This caching mechanism significantly reduces latency, particularly when models that share similar base tensors are loaded and saved frequently during iterative inference or fine-tuning.
It maintains a bounded size and is managed using a least-recently used (LRU) eviction policy.

\subsection{Delta Quantization Algorithm} \label{subsec:compressor}



\begin{figure}[t]
    \centering
    \includegraphics[width=0.86\linewidth]{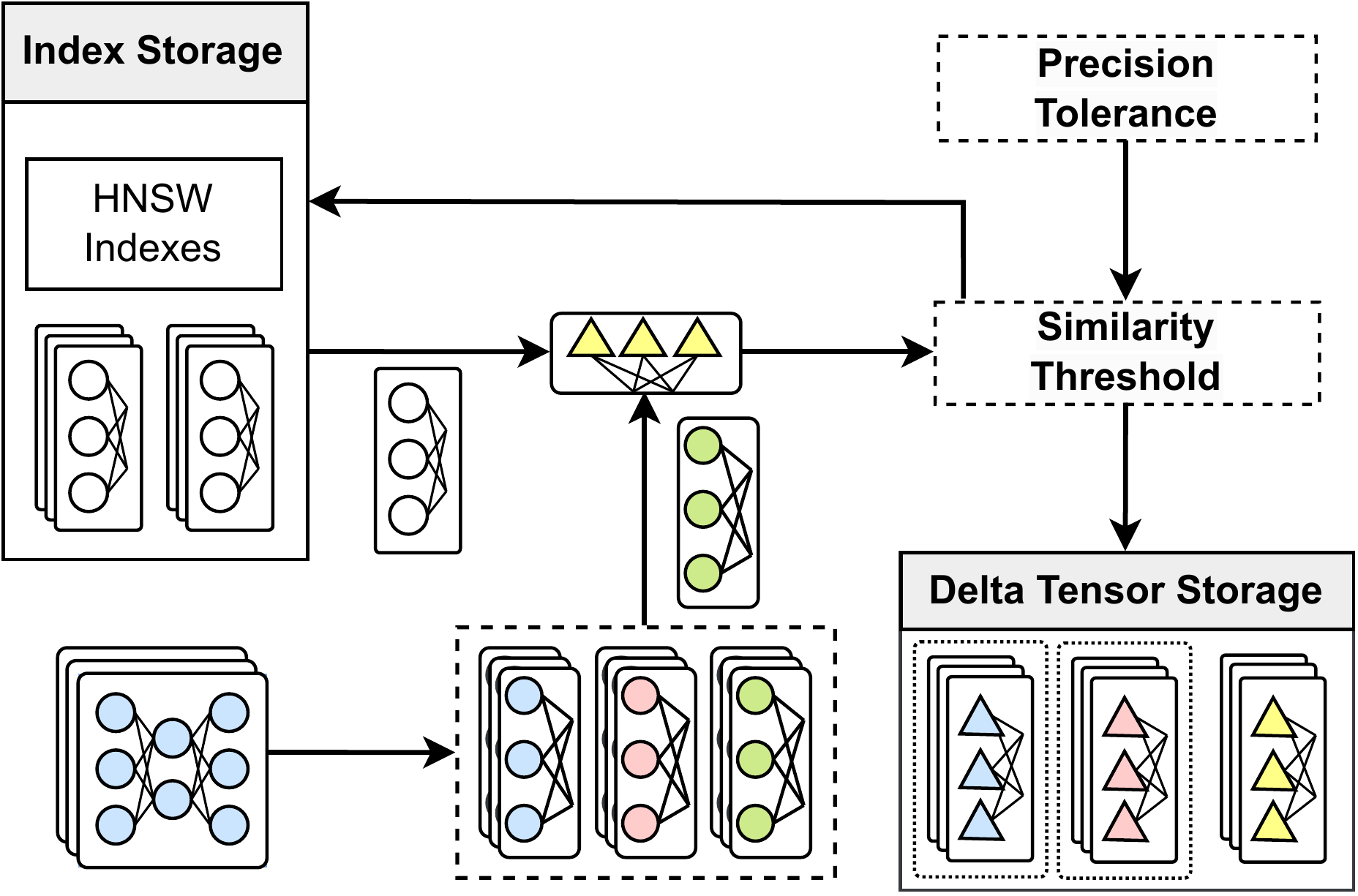}
    \vspace{-2mm}
    \caption{Delta Quantization Algorithm {\rm -- \dbname compresses tensors in four steps: (1) decouple the weights from model architecture, (2) search for the closest base tensor with ANN, (3) perform delta-encoding, and (4) apply quantization to deltas. }}
    \label{fig:delta_based_quantization}
\end{figure}

Based on the properties discussed in Section~\ref{subsec:motivation}, we propose a delta quantization algorithm to achieve efficient model compression while maintaining user-defined precision loss.
Figure~\ref{fig:delta_based_quantization} depicts the workflow of model insertion.
Given a collection of deep learning models $D=\{M_1,M_2,...,M_n\}$, and a user-defined precision tolerance $p$, \dbname compresses the model as follows.

\paragraph{Weight-Architecture Decoupling} Upon receiving the model saving request, we first decouple model weights (i.e., tensors) from model architectures to simplify and streamline the compression workflow.
In \dbname, deep learning model architectures and tensors are managed independently.
Given a set of deep learning models $D$, we extract their architectures into a set $S=\{s_1,s_2,...,s_n\}$, and aggregate all tensors into a unified set $T=\{t_1,t_2,...,t_m\}$, where $m$ is the total number of tensors across all models.
The architecture set $S$ is stored in full, while compression is applied solely to the tensor set $T$.
The benefit of such decoupling is that it enables independent optimization of the storage of tensors and architectures.
In particular, we can flatten the tensors to one-dimensional arrays, so that they can match with more similar tensors for deduplication.
For example, tensors with dimensions $(10, 10)$ and $(5, 20)$ are both flattened to a common shape of $(100, 1)$, increasing the opportunities of finding similar tensors.

\paragraph{Tensor Similarity Search} For each tensor $t \in T$, we search for a similar tensor already stored in \dbname using an approximate nearest neighbor index constructed for its shape. Specifically, we query the ANN index, $A$, to find a previously stored base tensor $t_{base}$ that minimizes the similarity metric (e.g., Euclidean distance) with $t$. In our design, we use the HNSW index for its efficiency and strong performance in searching high-dimensional tensors, which are prevalent in deep learning models.
As described in Section~\ref{subsec:storage}, the base tensor $t_{base}$ stored in the HNSW vertex consists of quantized 8-bit integers.
To enable the comparison with the input tensor, a 32-bit float, we de-quantize the base tensor, with the zero point and scale factor stored in the vertex, to $t_{full-base}$.

\paragraph{Delta Encoding}
We calculate the delta encoding of the tensor, $\delta$, and its bit width after quantization, denoted as $nbit$, as follows:
\begin{align}
    \delta & = t - t_{\textrm full-base}, \nonumber\\
    nbit & = \left\lceil \log_2\left(\frac{\delta_{max} - \delta_{min}}{2p}\right) \right\rceil. \label{eq:nbit}
\end{align}
where $\delta_{min}$ and $\delta_{max}$ are the minimum and maximum values in $\delta$, respectively.
We can observe that the range of $\delta$ and user-defined precision tolerance $p$ collaboratively determine the final bit width to store the tensor.
With a wide range of $\delta$ and a small precision tolerance $p$, the system will result in a large bit width and increase the storage consumption.
Therefore, we introduce a threshold $\tau$ for the range of $\delta$.
If $\delta_{max} - \delta_{min}$ is less than or equal to $\tau$, \dbname will proceed to quantize and store $\delta$ in the storage layer.
Otherwise, \dbname creates a new vertex in HNSW and recalculates the delta based on the new vertex.
This is to reduce the storage for the current tensor as well as potentially facilitate more effective compression for the future.
The procedure is detailed as follows:
(1) The original tensor $t$ is quantized to obtain $t_{quantized}$, comprising 8-bit integers.
(2) A new vertex storing $t_{quantized}$ is created and inserted into HNSW.
(3) The system applies de-quantization to $t_{quantized}$ to get $t'$. The delta is calculated using $\delta = t-t'$.
We have evaluated the impact of varying thresholds $\tau$ in Section~\ref{subsec:eval_sim_threshold}.
For a tensor with normalization, the range of the new delta falls within $\frac{1-(-1)}{2^8} \approx 0.0078$.
According to our evaluation, it is recommended to set a threshold between 0.1 and 0.2 to achieve the best performance.

\paragraph{N-Bit Quantization}
For tensors selected for compression, we quantize their delta values according to $nbit$ (calculated from Equation~\ref{eq:nbit}) to reduce storage cost while maintaining the precision loss within a user-defined precision tolerance $p$.
To achieve this, we apply linear asymmetric quantization with a fixed bit width of $2p$, setting the scale accordingly:
\begin{equation}
    quantized\_delta_i = \left\lfloor \frac{\delta_i}{scale} \right\rfloor + zero\_point,
\end{equation}
where $scale = 2p$, and $zero\_point = \left\lfloor -\frac{\delta_{min}}{scale} \right\rfloor$.
This indicates that the distance between two consecutive quantized numbers is $2p$, and therefore any points in between are within the distance of $p$ to their closest quantized number.
Moreover, since each tensor is individually quantized according to its value range, the number of bits required to store the tensor is minimized.



Algorithm~\ref{alg:model_compression} illustrates the complete model compression procedure.
Given a set of deep learning models $D = \{M_1, M_2, \ldots, M_n\}$ and a user-defined precision tolerance $p$, \dbname first decouples each model into its architecture and tensors (Line 1). The set of architectures $S$ is stored in full, while tensors $T$ are compressed.
For each tensor $t \in T$, the system performs an ANN search using an HNSW index to find the most similar base tensor $t_{\text{base}}$ (Lines 3-4). If a sufficiently similar match is found, the system computes the delta $\delta = t-t_{\text{base}}$ (Line 5). Otherwise, $t$ is quantized and inserted into the index, and the delta is recomputed against its quantized version (Lines 6-9). 
The delta tensor $\delta$ is then quantized using linear asymmetric quantization. The number of bits ($nbit$) is dynamically determined to ensure the quantization error remains within the tolerance $p$ (Line 10). Each element of $\delta$ is quantized using the derived scale and zero-point (Lines 11-14). The quantized delta, along with the reference tensor and bit width metadata, is stored for future reconstruction (Line 15).
Finally, the set of model architectures $S$ is saved to complete the compression process 
(Line 16).

\begin{algorithm}
    \small
    \DontPrintSemicolon
    \caption{Delta Quantization in \dbname}
    \label{alg:model_compression}

    \KwIn{Model set $D = \{M_1, M_2, \ldots, M_n\}$, precision tolerance $p$}
    \KwOut{Compressed tensors and model architectures stored in \dbname}

    $S, T \gets \textsc{DecoupleTensors}(D)$ \tcp*{Extract architectures $S$ and tensors $T$}

    \ForEach{$t \in T$}{
        $A \gets \textsc{GetIndexerFromPool}(dim(t))$\;
        $t_{\text{base}} \gets A.\textsc{Search}(t)$\;
        $\delta \gets \textsc{DeltaEncode}(t, t_{\text{base}})$\;

        \If{\textsc{ShouldCompress}($\delta$) = \textbf{false}}{
            $t_q \gets \textsc{QuantizeForIndex}(t)$\;
            $A.\textsc{Insert}(t_q)$\;
            $\delta \gets \textsc{DeltaEncode}(t, t_q)$\;
        }

        $nbit \gets \left\lceil \log_2\left( \frac{\max(\delta) - \min(\delta)}{2p} \right) \right\rceil$\;
        $\text{scale} \gets 2p$\;
        $\text{zero\_point} \gets \left\lfloor -\frac{\min(\delta)}{\text{scale}} \right\rfloor$\;

        \ForEach{$\delta_i \in \delta$}{
            $\text{qd}_i \gets \left\lfloor \frac{\delta_i}{\text{scale}} \right\rfloor + \text{zero\_point}$\;
        }

        \textsc{StoreQuantizedDelta}($\text{qd}, t_{\text{base}}, nbit$)\;
    }
    \textsc{StoreArchitectures}($S$)\;
\end{algorithm}

The delta quantization algorithm has three key novelties compared to existing methods.
First, it exploits inter-model similarities. For each new model inserted, we search globally for the tensors closest to the input tensor, resulting in deduplication among models.
Moreover, the integration of ANN enables flexible and incremental compression.
As the system ingests more models, the growing number of base tensors increases the likelihood of finding close matches for newly added tensors, thereby improving compression efficiency over time.
This approach eliminates the need to re-compress existing models and is particularly well-suited for dynamic model management scenarios where models are frequently added and fine-tuned.
Second, it applies quantization to delta encoding. 
Compared with quantization over the original weights, this method substantially reduces the required bit width to represent model weights and lowers the precision loss by processing on a smaller scale.
Lastly, each tensor and delta is dynamically quantized based on its value range, rather than applying a fixed global quantization parameter.
This adaptive approach maximizes the compression ratio while adhering to the user-defined precision loss constraints.

\paragraph{
Discussion
}
The effectiveness of the delta quantization algorithm is decided by the precision tolerance $p$, which defines the upper bound of the quantization bin width. A larger precision tolerance results in quantization with wider rounding intervals, thus yielding a higher compression ratio, but potentially degrading model performance. To mitigate such risk, \dbname uses a precision tolerance of $5.96 \times 10^{-8}$ ($2^{-24}$) by default, which is smaller than the machine epsilon for single-precision floating-point numbers. 
As shown in Section~\ref{subsec:eval_model_performance}, over 90\% of the tested models exhibit no performance change under this tolerance, demonstrating that the default precision tolerance is sufficiently strict to limit the impact on model performance.
\dbname also allows users to configure the precision tolerance on a per-model basis. 
We provide a utility tool in our code repository~\cite{neuralstoreimpl} to guide users in selecting an appropriate tolerance for a specific model.
First, given a model and a test dataset, it compresses the model using multiple candidate tolerances.
Next, it evaluates the performance of each compressed variant on the test data and reports the results to the user. 
Based on the analysis results, users can choose a preferred tolerance to store the model accordingly, effectively balancing storage consumption and model performance degradation.

\subsection{Compression-aware Model Loading}

\begin{figure}
    \centering
    \includegraphics[width=0.8\linewidth]{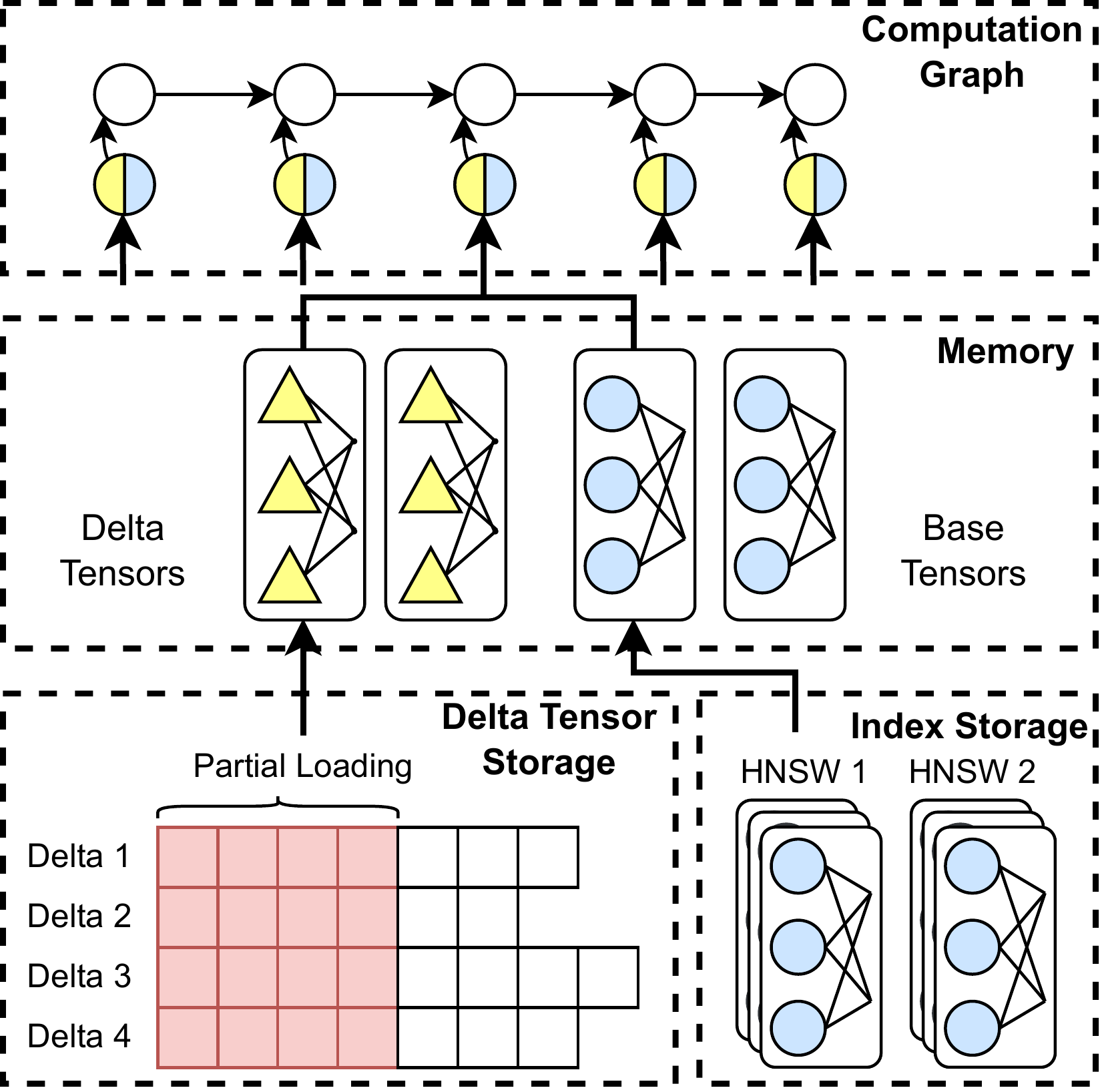}
    \vspace{-2mm}
    \caption{Compression-aware Model Inference {\rm -- \dbname adopts flexible tensor loading with partial delta tensor bits and on-demand decompression to streamline the model loading process.}}
    \label{fig:loading}
\end{figure}

In conventional model management systems, compressed models are often required to be decompressed or reconstructed before they are served for inferences.
This results in significant overhead in model loading and extensive memory consumption, accommodating the full model inside memory.
As shown in Figure~\ref{fig:loading}, we present the compression-aware model inference mechanism, which streamlines the model loading and inference process, as well as reduces the memory consumption.

\subsubsection{Model Loading}
When a {\it Load Model} request is received, \dbname first looks up the reference of the first tensor page in the model table with the model ID. Since the tensor pages of a model are organized consecutively, it then scans the delta pages for model architecture, delta tensors, and references (HNSW ID and vertex ID) to base tensors. 
Lastly, \dbname traverses the HNSWs to fetch the index pages containing base tensors.
\dbname only stores the quantized tensors in memory to reduce the memory consumption.

\paragraph{Flexible Model Loading}
\label{sec:flexibl_model_loading}
\dbname enables flexible model loading to optimize the trade-off among memory consumption, loading efficiency, and precision loss.
In scenarios where the efficiency of model serving is critical, while higher model tolerance is accepted, \dbname allows the users to selectively fetch partial bits of delta tensors, or even only the base tensor.
This will lead to faster model loading and lower memory consumption due to fewer bits loaded and disk I/O at the cost of higher model precision loss.
Notably, the additional precision loss brought by the flexible model loading only has a limited impact on the resulting model performance due to our unique compression algorithm, as shown in Section~\ref{sec:evaluation}.
Since each delta is calculated with respect to the closest base tensor, and quantization is applied dynamically on each delta, ignoring the least significant bits of the quantized delta leads to an average difference of $10^{-4}$ compared with fetching the quantized delta in full bits.

\subsubsection{On-demand Decompression}
\label{sec:on_demand_decompression}
\dbname adopts the on-demand decompression during model serving to ensure the minimum memory usage.
To serve the model, \dbname first deserializes the model architecture to form a computation graph.
When the computation reaches the step that involves a compressed tensor, it will de-quantize the delta tensor and the corresponding base tensor and reconstruct them to get the full-bit-width tensor for calculation.
The full-bit-width tensor will be discarded after the computation is finished.
This leads to consistent memory usage with additional computation cost for decompression.
Such overhead can be mitigated by temporarily storing the de-quantized base tensors, which will be used again by the following layers.
In particular, during model loading, we record the share count of each base tensor.
For base tensors with the share count greater than 0 during the computation, we store the de-quantized base tensor and decrease the share count.
When the share count reaches 0, the de-quantized base tensor will be deleted.
In this way, we eliminate the duplicate de-quantization of the same base tensors.

\paragraph{Augmented Computation Graph}
To enable on-demand decompression inference, \dbname augments the original model graph with additional computational nodes that handle dequantization and reconstruction at runtime. Specifically, a compressed tensor is reconstructed by combining two components: the quantized base tensor and its corresponding quantized delta. The base tensor is always stored and loaded in 8-bit quantized form, while the delta tensor is flexibly loaded based on the desired inference precision, as discussed in Section~\ref{sec:flexibl_model_loading}.
Both the base and delta tensors are dequantized using their associated scale and zero-point values, which are retrieved alongside the quantized representations. These dequantization operations are expressed as \texttt{DequantizeLinear} nodes within the computation graph. The outputs of the two dequantization branches are then combined through an element-wise addition node to reconstruct the original tensor. This augmented graph eliminates the need for full offline decompression and enables efficient execution directly over the compressed representation.

\begin{algorithm}[t]
    \small
    \DontPrintSemicolon
    \caption{Compression-Aware Model Inference}
    \label{alg:model_inference}

    \KwIn{Compressed model $\hat{M}$, inference bit width $b$}
    \KwOut{Augmented computation graph $G$ for runtime execution}

    $G \gets \textsc{LoadModelGraph}(\hat{M})$\;
    $T \gets \textsc{GetCompressedTensors}(\hat{M})$\;

    \ForEach{tensor $t \in T$}{
        $t_{\text{base}} \gets \textsc{RetrieveQuantizedBase}(t)$\;
        $t_{\text{delta}}, nbit \gets \textsc{RetrieveQuantizedDelta}(t)$\;

        \If{$nbit > b$}{
            $t_{\text{delta}} \gets \textsc{ExtractMSB}(t_{\text{delta}}, b)$\;
            $\text{scale}_{\text{delta}} \gets \text{scale}_{\text{delta}} \times 2^{nbit - b}$\;
        }

        $N_{\text{deq\_base}} \gets \textsc{CreateDequantizeNode}(t_{\text{base}}, \text{scale}_{\text{base}}, \text{zp}_{\text{base}})$\;
        $N_{\text{deq\_delta}} \gets \textsc{CreateDequantizeNode}(t_{\text{delta}}, \text{scale}_{\text{delta}}, \text{zp}_{\text{delta}})$\;

        $N_{\text{add}} \gets \textsc{CreateAddNode}(N_{\text{deq\_base}}, N_{\text{deq\_delta}})$\;
        $G.\textsc{InsertNode}(N_{\text{add}})$\;
        $G.\textsc{DirectOutput}(N_{\text{add}}, \textsc{OriginalNode}(t))$\;
    }

    \Return{$G$}
\end{algorithm}

We illustrate the compression-aware model inference process in Algorithm~\ref{alg:model_inference}.
Given a compressed model $\hat{M}$ and the targeted delta inference bit width $b$, \dbname first loads the original computation graph (Line 1).
For each tensor, the system retrieves its quantized base and delta components, along with their associated quantization metadata (i.e., bit width, scale, and zero-point) (Lines 4-5).
If the delta tensor was originally quantized to a higher bit width than the target delta bit width $b$, only the most significant $b$ bits are extracted. To compensate for the bit truncation, the quantization scale is adjusted proportionally (Lines 6-8).
The system then creates two \texttt{DequantizeLinear} nodes, one for the base tensor and one for the (truncated) delta tensor (Lines 9-10).
These two dequantized outputs are fused through an \texttt{Add} node to reconstruct the tensor in float space (Line 11).
\dbname inserts the dequantization and addition nodes directly into the model graph. The output of this reconstruction subgraph is then wired to the corresponding original node that consumed the tensor (Lines 11-13). This augmentation enables runtime reconstruction of compressed tensors and eliminates the need for full offline decompression.

\subsubsection{Pipelining}
\begin{figure}
    \centering
    \includegraphics[width=0.75\linewidth]{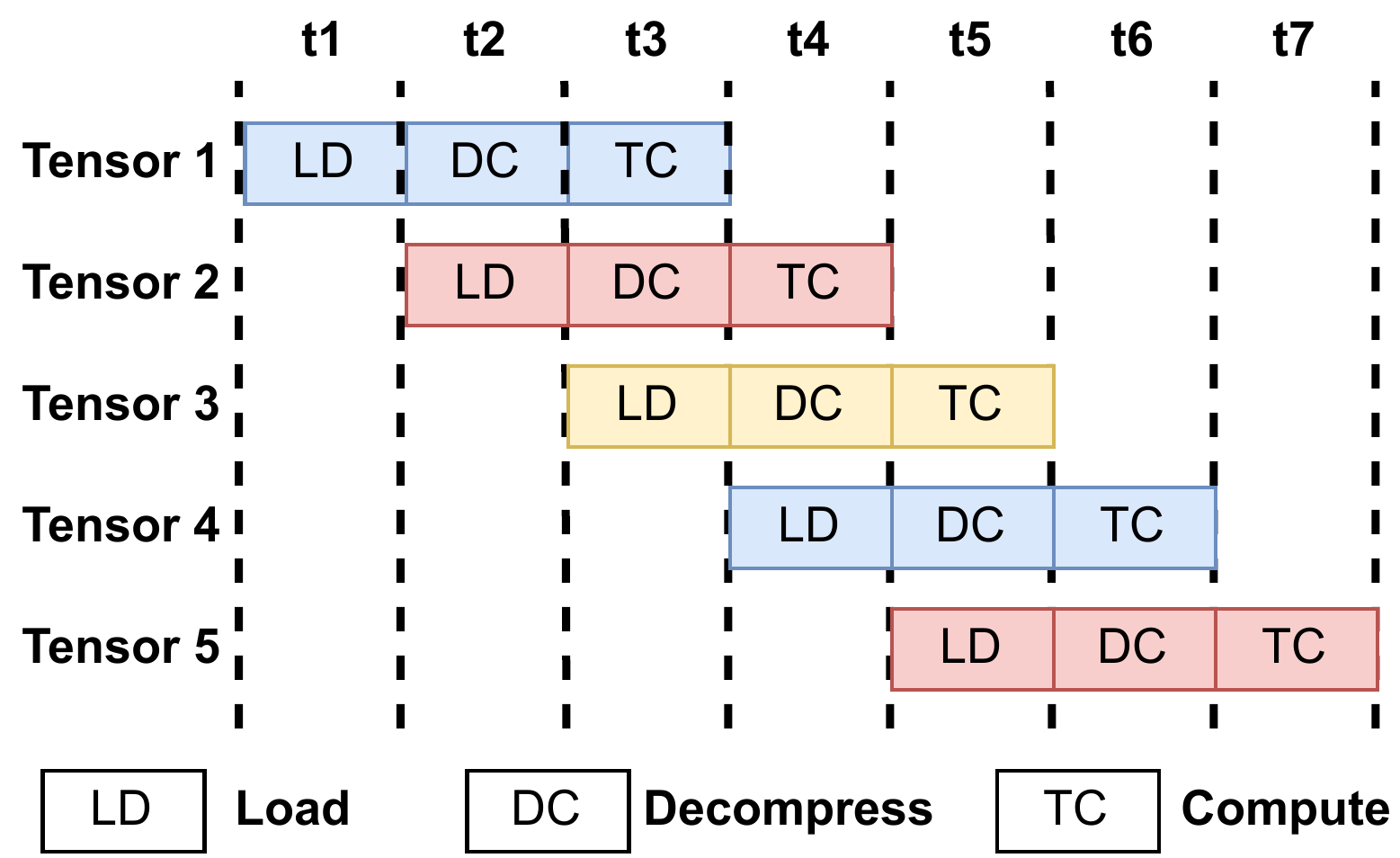}
    \vspace{-2mm}
    \caption{Pipelining {\rm -- \dbname pipelines tensor loading, decompression, and computation during model loading.}}
    \label{fig:pipline}
\end{figure}

We further improve the model serving process by leveraging pipelining.
The entire process can be divided into three phases, namely, model loading, tensor decompression, and model computation. 
The model loading phase is I/O-intensive, as it involves retrieving delta tensors and HNSW indices that store the base tensors.
Tensor decompression is primarily CPU-bound, though its computational overhead can be mitigated through the use of AVX instruction sets.
Model computation is also CPU-intensive, dominated by matrix multiplication operations.
To improve throughput, we pipeline these three phases. As illustrated in Figure~\ref{fig:pipline}, at the $i$-th stage of the pipeline, the system performs model loading for the $i$-th tensor, decompression for the $(i-1)$-th tensor, and model computation for the $(i-2)$-th tensor. This design enables concurrent execution of the three phases, effectively hiding latency and improving resource utilization.
The benefits of pipelining are further amplified when the model computation phase is offloaded to a GPU due to less resource contention.

\section{Implementation} \label{sec:implementation}

We implement \dbname as a pluggable extension of PostgreSQL with 5000 LoC lines of code in C/C++. 
We have made our source code available~\cite{neuralstoreimpl}.

\dbname is tightly integrated with PostgreSQL to support efficient model storage, loading, and compression-aware inference through SQL interfaces.
We provide several PostgreSQL UDFs, such as \texttt{ns\_save\_model} and \texttt{ns\_load\_model}, to store and retrieve models. 
\dbname manages model storage at the granularity of tensors. Compressed tensors are organized into \emph{tensor pages}, each of which contains the complete set of compressed tensors for a single deep learning model. Tensor pages are enforced to be read-only by \dbname, and are memory-mapped using \texttt{mmap} to reduce memory footprint and enable sharing across PostgreSQL sessions. At the start of each tensor page, a fixed-length header records the offsets and lengths of all delta tensors. Each delta tensor keeps metadata, including its shape, dimension, quantization parameters (i.e., scale and zero point), and single-element bit width, followed by a bit-packed payload.

To accelerate the retrieval of HNSW indexes, \dbname maintains a local cache for HNSW indexes. 
The cache is bounded by a user-defined buffer size, which is set to 32 GB by default at the startup stage of PostgreSQL. 
When memory is insufficient, the least recently used index is evicted from the cache to the disk based on LRU.
We build our index on top of hnswlib, and extend a new \texttt{SpaceInterface} called \texttt{QuantizedL2Space} to support distance computation between quantized vectors with different scales and zero-points. To further accelerate this, we optimize the distance computation using AVX2 SIMD instructions.

\dbname supports both full decompression and compression-aware inference for ONNX framework models. For compression-aware inference, we modify ONNX computation graphs to incorporate on-demand decompression. Particularly, the system retrieves quantized delta and base tensors and inserts \texttt{DequantizeLinear} and \texttt{Add} nodes into the graph to perform runtime reconstruction, as described in Section~\ref{sec:on_demand_decompression}. During compression, both delta computation and quantization are carried out in double precision to mitigate rounding errors introduced by low-precision representations.



To demonstrate the generalizability of the proposed method, we also implement \dbname as a loadable extension of DuckDB. 
Adapting \dbname to DuckDB involves three customizations: (1) rewriting the UDF registration logic, (2) replacing PostgreSQL's shared-memory-based index cache with an in-process index cache, and (3) adapting memory allocation from PostgreSQL to DuckDB's C++ runtime.
This DuckDB-based implementation is functionally comparable to the PostgreSQL-based implementation, enabled by two key design factors.
First, \dbname relies on components that are commonly available in modern DBMSs, such as page-based storage engines, UDF interfaces, and buffer managers. 
Second, its modular design decouples database-specific APIs, such as shared memory management and data access methods, from the core logic of model compression and loading.

\section{Performance Evaluation} \label{sec:evaluation}

In this section,
we evaluate \dbname by comparing it with state-of-the-art model management systems and compression algorithms.
We conduct system-level and micro benchmarks to measure the query throughput, storage consumption, and model accuracy.

\subsection{Experimental Setup}
\label{sec:exp:settings}

We conduct our experiments on two servers, each equipped with an Intel(R) Xeon(R) W-1290P CPU (10 cores $\times$ 2 hardware threads), 128GB of DRAM, a 894GB SAMSUNG\_MZ7L3960 SSD, and an NVIDIA RTX 3090 GPU.
We use one server to simulate clients that issue save model and load model requests, while the other server functions as the database server.

\subsubsection{Workloads} \label{subsec:workloads}

To evaluate model management performance under realistic scenarios, we collect 800 DL models totaling 361GB, covering a diverse range of model architectures and sizes.
Before running the experiments, we perform a warmup phase to download all models from Huggingface~\cite{huggingface} and store them locally in advance.
We feed the systems with read and write workloads, containing randomly selected models, to simulate model saving and loading operations.
We first save the models into the evaluated system to measure write throughput and subsequently load them to measure read throughput.


As summarized in Table~\ref{tab:workloads}, we set up three representative in-database AI-powered analytics tasks: 
(a) sequence classification~\cite{DBLP:journals/pvldb/BruckeHPPR23}: 20 DistilBERT models fine-tuned on the IMDB dataset~\cite{dataset_imdb}, each performing 100 text inferences to classify movie reviews, 
(b) image classification~\cite{EVA_SIGMOD22}: 20 Vision Transformer (ViT) models fine-tuned on the Beans dataset~\cite{dataset_beans}, each processing 100 images to identify bean leaf diseases, and 
(c) tabular classification~\cite{DBLP:journals/corr/abs-2408-03013}: 12 multi-layer perceptron (MLP) models trained on the Avazu dataset~\cite{dataset_avazu}, each classifying 500 user records to predict CTR scores.

\begin{table}[t]
    \centering
    \caption{
    Experimental AI-powered Analytics Workloads
    }
    \label{tab:workloads}
    \vspace{-2mm}
    \resizebox{0.9\columnwidth}{!}{%
    \begin{tabular}{cccc}
        \hline
        \textbf{ID} & \textbf{Workload} & \textbf{DataSet} & \textbf{Model} \\
        \hline
        (a) & Sequence Classification & IMDB & DistilBERT \\
        (b) & Image Classification & Beans & Vision Transformer \\
        (c) & Tabular Classification & Avazu & MLP \\
        \hline
    \end{tabular}
    }
\end{table}

\extended{
\begin{table}[t]
    \centering
    \caption{Summary of models used in the workloads}
    \label{tab:workloads}
    \vspace{-2mm}
    \resizebox{0.7\columnwidth}{!}{%
    \begin{tabular}{cccc}
        \hline
        \textbf{Type} & \textbf{Model} & \textbf{Count} & \textbf{Size (GB)} \\
        \hline
        \multirow{6}{*}{CV} & MobileNetV2 & 30 & 0.41 \\
        & ResNet-50 & 110 & 10.53 \\
        & ViT-B/16 & 110 & 35.45 \\
        & ViT-L/32 & 5 & 5.85 \\
        & Swin-T & 70 & 7.52 \\
        & Swin-B & 60 & 20.51 \\
        \hline
        \multirow{8}{*}{NLP} & BERT-base & 50 & 22.15 \\
        & DistilBERT & 50 & 12.21 \\
        & RoBERTa & 50 & 24.41 \\
        & BERT-large & 50 & 58.59 \\
        & T5-small & 50 & 11.72 \\
        & T5-base & 50 & 43.46 \\
        & BART-base & 50 & 21.48 \\
        & BART-large & 50 & 78.13 \\
        \hline
        Multimodal & BLIP-base & 15 & 5.86 \\
        \hline
        \textbf{Total} & & \textbf{800} & \textbf{358.28} \\
        \hline
    \end{tabular}
    }
    \vspace{-4mm}
\end{table}
}

\subsubsection{Baselines}

We compare it against two representative baseline systems: a database-based model management system, PostgresML, and a file-based model management system, ELF$^\ast$.
To facilitate fair comparisons, all systems are integrated with PostgreSQL to store the metadata of models.
We employ an open-sourced implementation on PostgreSQL~\cite{DBLP:journals/corr/abs-2408-03013} that provides standard in-database AI-powered analytics interfaces, and integrate it with the evaluated systems to perform end-to-end experiments, i.e., conducting model saving, model loading, and model inference within the database.


\paragraph{PostgresML~\cite{postgresml}}
PostgresML is a PostgreSQL extension designed for in-database machine learning.
It stores each model as a serialized BLOB in a model table and leverages PostgreSQL's built-in TOAST mechanism for compression, using a LZ-family compression algorithm known as PGLZ.

\paragraph{ELF$^\ast$~\cite{comp_Elves_VLDB2024}}
To the best of our knowledge, there is no open-source file-based model management system.
We therefore implement ELF$^\ast$, a file-based baseline that integrates ELF~\cite{comp_Elves_VLDB2024}, a state-of-the-art model compression algorithm.
When saving new models, ELF$^\ast$ compresses each model using ELF and stores it as a separate file, and then records the file path in a PostgreSQL model table.
When loading a model, ELF$^\ast$ fetches the model metadata (including the file path) from the database, loads the model file from disk, and decompresses it for use.


We also compare \dbname with widely used general-purpose compression algorithms, ZSTD and ZFP, and specialized model compression methods, ELF, in terms of storage consumption and the resulting model accuracy.

\paragraph{ZSTD~\cite{zstandard}}
ZSTD is a lossless compression algorithm developed by Facebook. It is commonly used for general-purpose data compression.
We use its official release (v1.5.5) in our experiments.

\paragraph{ZFP~\cite{zfp}}
ZFP is a lossy compression algorithm designed for floating-point arrays.
It allows users to adjust error bounds according to the accuracy requirements of the target data.
We use its official release (v0.5.5) to conduct our experiments.

\paragraph{ELF~\cite{comp_Elves_VLDB2024}}
ELF is a state-of-the-art model compression framework that eliminates the exponent fields of floating point numbers by projecting the model weights from (-1,1) to [1, 2).
We use its official open-source code and extend it to support the ONNX model format to align with our workload.

\subsubsection{Default configuration}
We configure PostgreSQL with a 32GB shared buffer.
Other PostgreSQL parameters remain at their default values unless specified.
We choose $\tau=0.16$ as the default similarity threshold in \dbname, and enable the flexible model loading by default.
We set the precision tolerance of ZFP and \dbname to $5.96 \times 10^{-8}$, consistent with that used in ELF, ensuring a uniform upper bound on the precision loss.
A detailed analysis of how these default parameter settings are determined is provided in Section~\ref{subsec:eval_micro}.

\subsection{System Performance Evaluation}

Here we evaluate the system performance of \dbname by comparing it with PostgresML and ELF$^\ast$.

\subsubsection{
End-to-end Performance Evaluation
}
\label{sec:eval_end_to_end}
\begin{figure}
\centering
\includegraphics[width=0.96\linewidth]{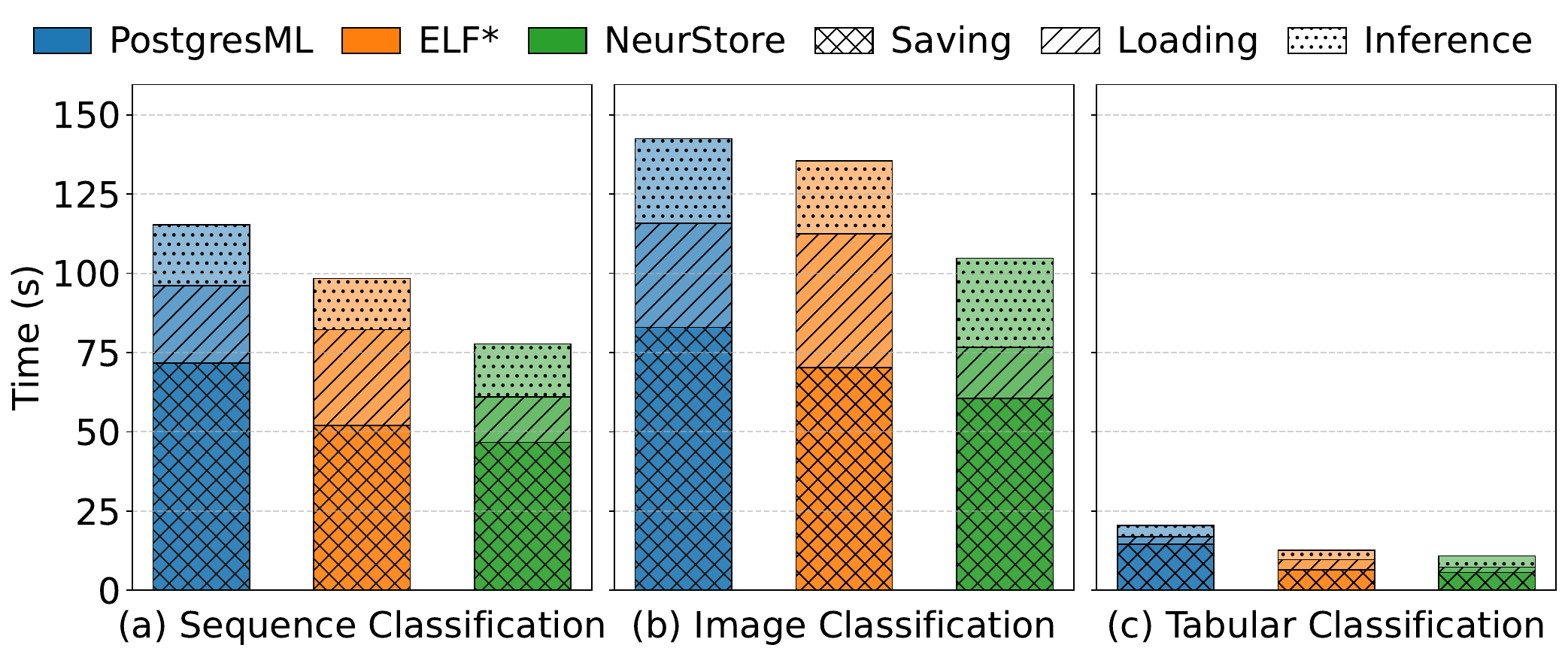}
\vspace{-4mm}
\caption{
End-to-end Time Breakdown for In-database AI-powered Analytics
}
\label{fig:end_to_end_breakdown}
\vspace{-4mm}
\end{figure}

We first evaluate the end-to-end latency for in-database AI-powered analytics. 
In each task, we save multiple models as described in Section~\ref{sec:exp:settings}, then load each model and perform inference.
We accumulate the latency of each stage and show the results in Figure~\ref{fig:end_to_end_breakdown}.
\dbname reduces total query latency by up to 32\%, 26\%, and 47\% compared to PostgresML, and by up to 20\%, 22\%, and 14\% compared to ELF$^\ast$, across sequence, image, and tabular classification tasks.
For model saving, \dbname takes 46s, 60s, and 5s for the three tasks, outperforming both PostgresML (71s, 82s, 14s) and ELF$^\ast$ (52s, 70s, 6s).
This improvement is attributed to our tensor-based storage engine, which batches tensor writes into pages and defers HNSW index updates until eviction, thereby reducing I/O overhead.
For model loading, \dbname achieves up to 61\% and 50\% speedup over ELF$^\ast$ and PostgresML, respectively.
This is due to the fact that our tensor-based storage engine avoids redundant base-tensor fetching and reduces disk I/O.
Moreover, our compression-aware model loading mechanism eliminates decompression during loading, further improving the model loading throughput.
Inference latency remains comparable across all systems since they share the same ONNX runtime, with only negligible increases for \dbname due to the on-demand decompression during inference.


\begin{figure}[t]
\centering
\scalebox{0.96}{\includegraphics[width=0.65\linewidth]{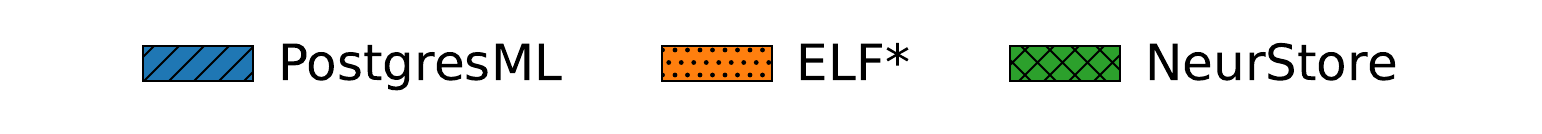}}
\par\vspace{-2mm}
\subfigure[Write Throughput]{
\label{Fig.indb_analytics.throughput}
\includegraphics[width=0.32\linewidth]{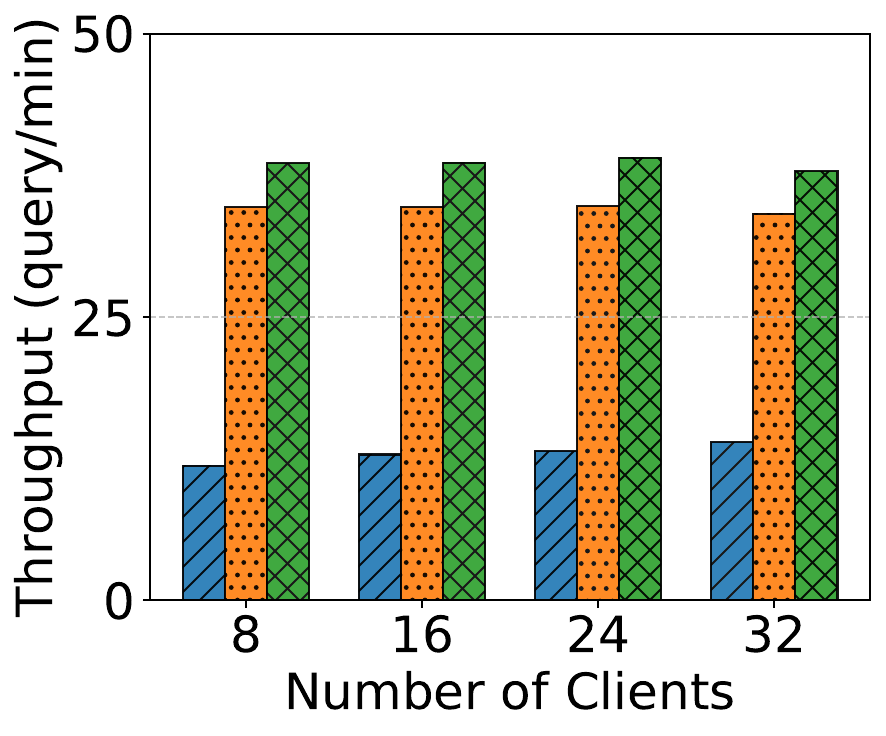}
}
\subfigure[Read Throughput]{
\label{Fig.indb_analytics.latency}
\includegraphics[width=0.32\linewidth]{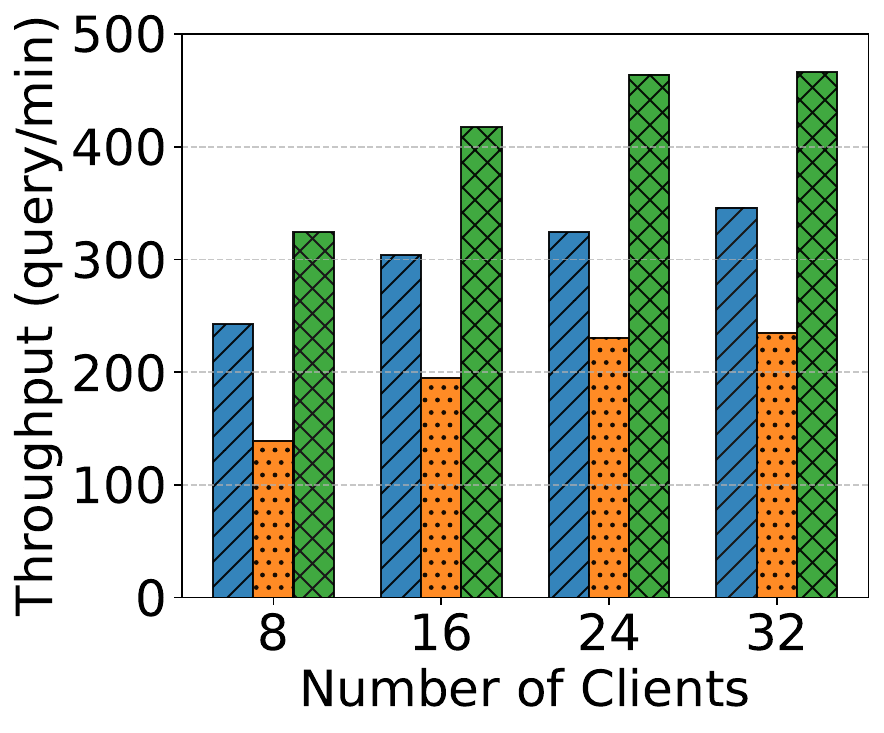}
}
\subfigure[Storage Consumption]{
\label{Fig.indb_analytics.storage}
\includegraphics[width=0.28\linewidth]{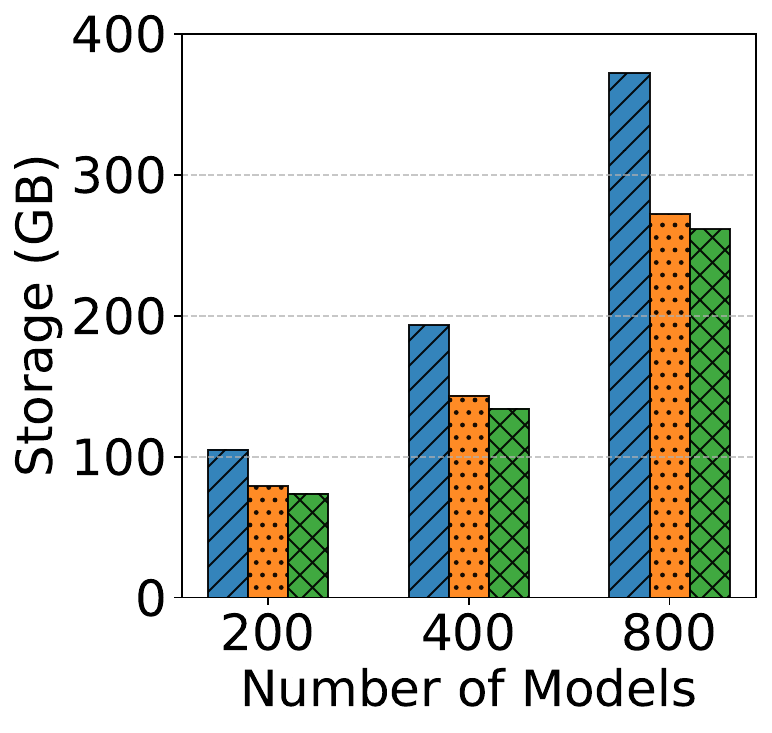}
}
\vspace{-4mm}
\caption{Overall Performance of \dbname.}
\label{fig.end_to_end_tps}
\end{figure}

\begin{figure}
\centering
\begin{minipage}{\linewidth}
    \centering
    \subfigure[Storage Consumption]{
        \includegraphics[width=0.47\linewidth]{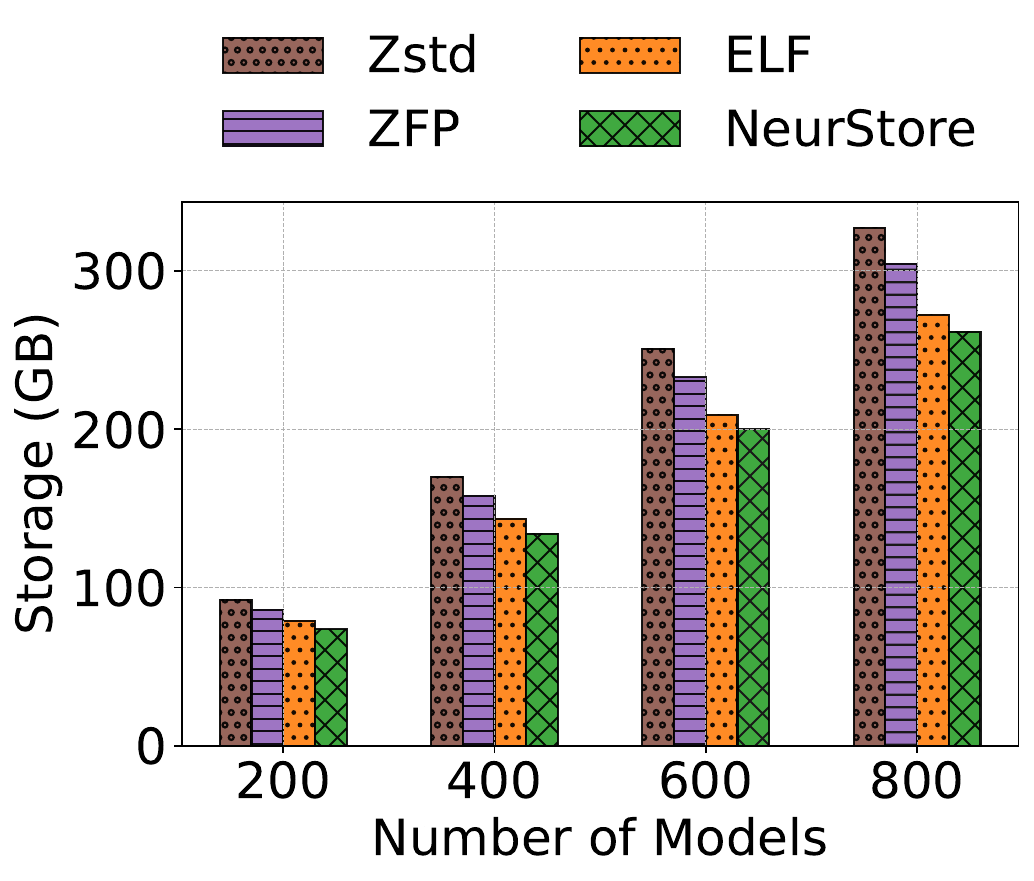}
    }
    \subfigure[Compression Ratio]{
    \label{fig:cdf_storage}
        \includegraphics[width=0.47\linewidth]{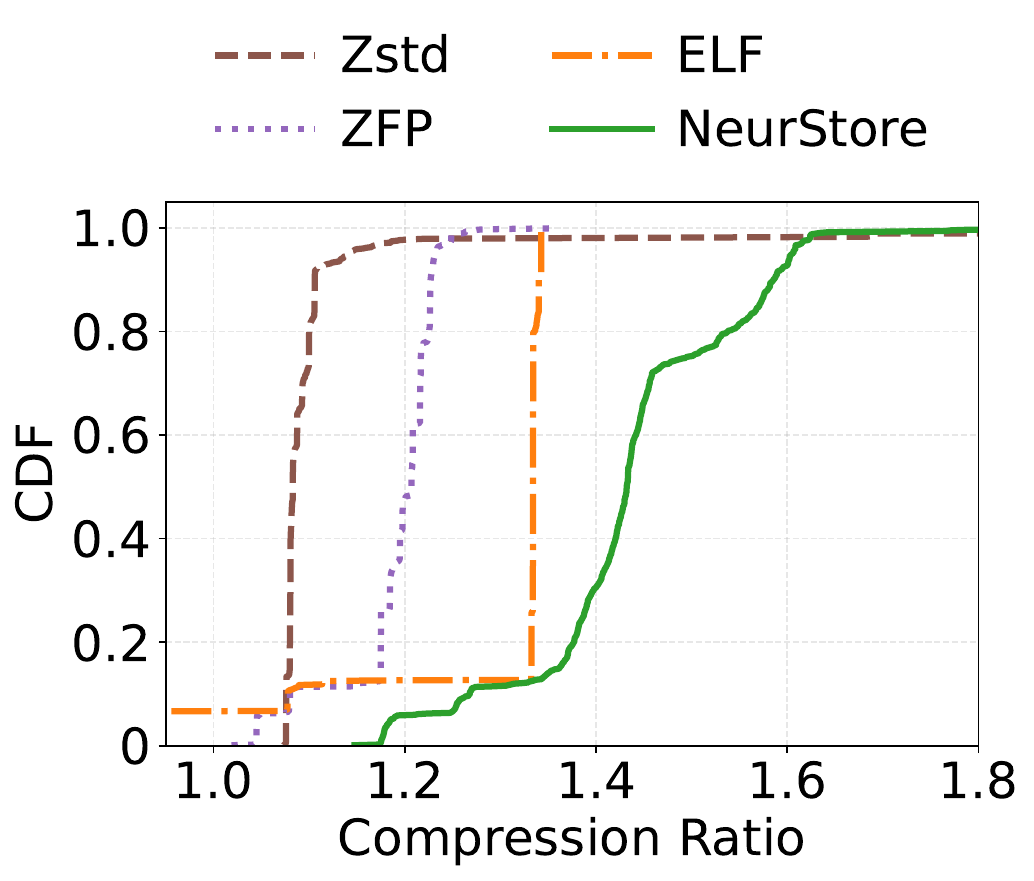}
    }
    \vspace{-4mm}
    \caption{Evaluation on Compression Algorithms.}
    \vspace{-4mm}
    \label{fig:eval_storage}
\end{minipage}
\end{figure}

\begin{figure}
\begin{minipage}{\linewidth}
    \centering
    \subfigure[Storage Consumption]{
        \label{fig:eval_tensor_based_storage_storage}
        \includegraphics[width=0.47\linewidth]{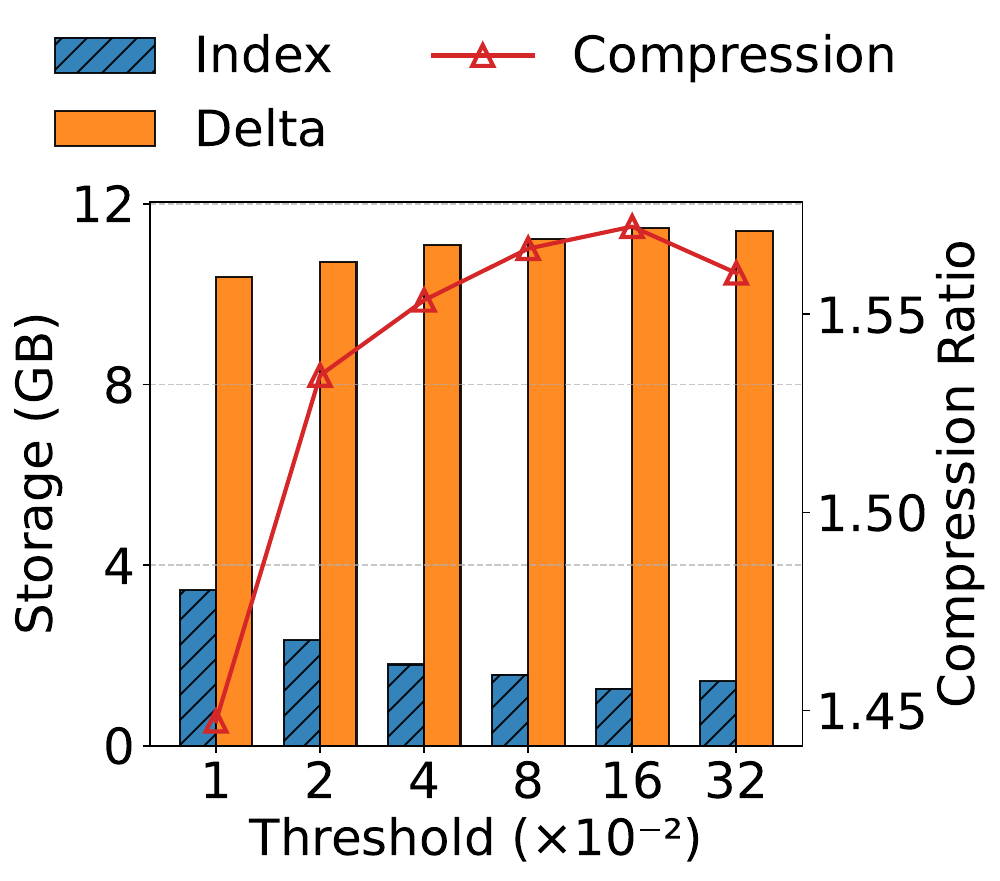}
    }
    \subfigure[Compression Throughput]{
        \label{fig:eval_tensor_based_storage_throughput}
        \includegraphics[width=0.47\linewidth]{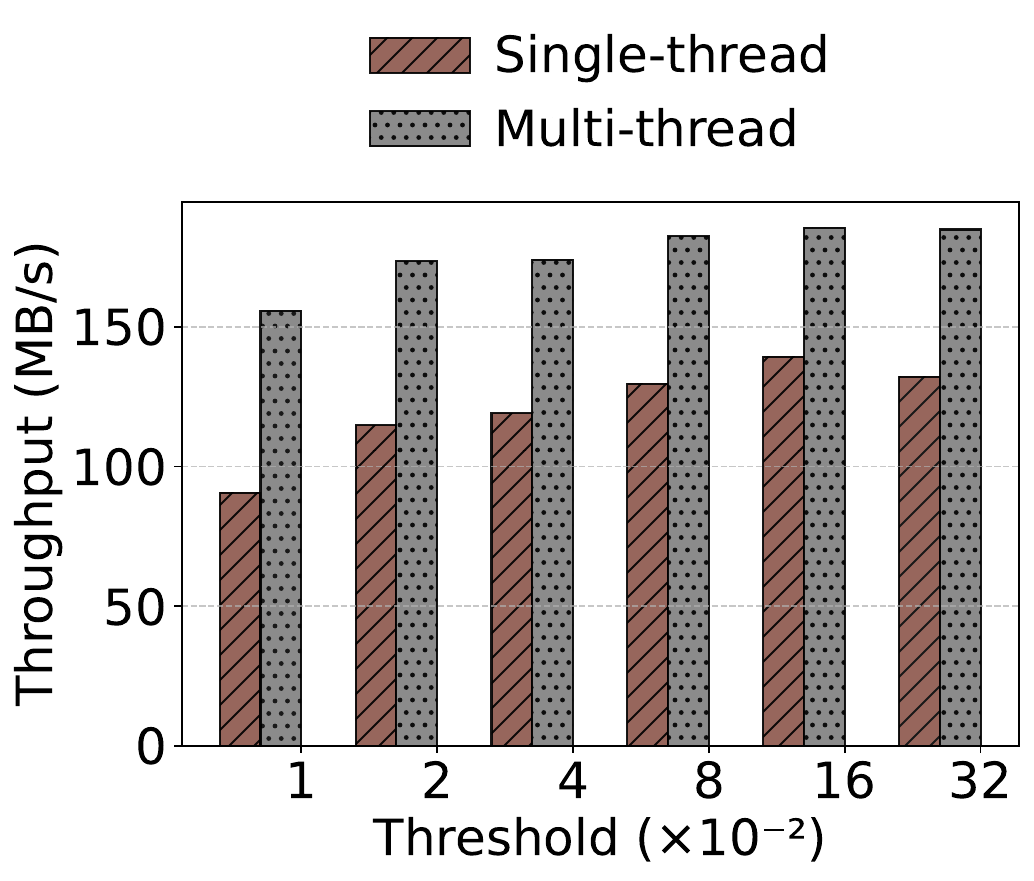}
    }
    \vspace{-4mm}
    \caption{Performance Impact of Similarity Threshold.}
    \label{fig:eval_tensor_based_storage}
\end{minipage}
\end{figure}

\subsubsection{Throughput.}
\label{sec:eval_throughoput}
We then evaluate the write and read throughput with varying numbers of clients ranging from 8 to 32.
For write operations, the clients concurrently save models randomly chosen from the model pool described in Section~\ref{subsec:workloads}.
The results are shown in Figure~\ref{Fig.indb_analytics.throughput}.
We can observe that the throughput of all systems increases as more clients are included due to the increased concurrency.
\dbname outperforms the baselines, achieving a peak throughput of 39 queries per minute compared with 35 for ELF$^\ast$ and 14 for PostgresML.
PostgresML performs the worst by storing the model with the TOAST mechanism in PostgreSQL.
The models are divided into small chunks, which are fetched separately and reconstructed.
This leads to high disk I/O overhead.
ELF$^\ast$ achieves throughput comparable to \dbname across all numbers of clients, but remains behind due to higher I/O cost, especially when multiple models are stored concurrently.
For read operations, each client continuously sends queries, with each query loading a random model.
Figure~\ref{Fig.indb_analytics.latency} shows the read throughput. 
\dbname outperforms ELF$^\ast$ and PostgresML by up to $2.5\times$ and $1.4\times$, respectively.
The improvement in the performance stems from three key designs. 
First, \dbname maintains an in-memory index cache to avoid repeated fetching of base tensors, thus reducing the disk I/O.
Second, \dbname adopts an on-demand decompression strategy, where quantized deltas and base tensors are de-quantized at inference time, thereby eliminating the need for full model decompression prior to execution.
Lastly, the flexible model loading enables \dbname to load only the most significant 8 bits of the quantized deltas for inference, further reducing the disk I/O and memory bandwidth.

\subsubsection{Storage.}
\label{subsec:eval_storage}
We report the resulting storage usage in Figure~\ref{Fig.indb_analytics.storage}.
As observed, \dbname has the least storage consumption.
In particular, it consumes 93\% and 70\% of the space required by ELF$^\ast$ and PostgresML, respectively.
Overall, \dbname achieves a compression ratio of 1.38$\times$, compared to 1.32$\times$ for ELF$^\ast$ and 0.97$\times$ for PostgresML.
These improvements are due to 
\dbname's
delta quantization compression, which identifies shared base tensors across models and dynamically quantizes the base and delta tensors to reduce storage cost.

\subsubsection{In-depth Bottleneck Analysis}
\label{subsec:eval_sys_prof}
To better understand system bottlenecks, we report CPU and I/O costs for saving and loading a representative model (google/vit-base-patch16-224) in Table~\ref{tab:statistics}.
For model saving, \dbname achieves the shortest wall time and lowest I/O block usage. This aligns with the results in Figure~\ref{Fig.indb_analytics.throughput}.
In addition, \dbname exhibits the highest CPU utilization compared to PostgreSQL and ELF$^\ast$, which is expected because its higher compression ratio incurs greater computational cost.
For model loading, \dbname achieves up to 50\% and 55\% lower wall time than PostgresML and ELF$^\ast$, respectively. The results are also consistent with its higher read throughput shown in Figure~\ref{Fig.indb_analytics.latency}.
We also observe that \dbname achieves the lowest system time, CPU utilization, and I/O block reads.
This is attributed to the on-demand decompression and flexible loading strategies, which reduce the cost of fully decompressing models and I/O overhead.
In addition, we measure the memory usage of the evaluated model.
The results show that \dbname consumes 165MB of memory during model loading, compared to 330MB for both PostgresML and ELF$^\ast$.
This further demonstrates the effectiveness of the proposed flexible loading mechanism in reducing memory consumption.

\begin{table}[t]
    \centering
    \caption{
    CPU and I/O Statistics for Model Saving / Loading
    }
    \label{tab:statistics}
    \vspace{-2mm}
    \resizebox{0.99\columnwidth}{!}{%
    \begin{tabular}{ccccccc}
        \hline
        \textbf{Operation} & \textbf{System} & \makecell{\textbf{Wall}\\\textbf{Time (s)}} & \makecell{\textbf{User}\\\textbf{Time (s)}} & \makecell{\textbf{System}\\\textbf{Time (s)}} & \makecell{\textbf{CPU}\\\textbf{Utilization}} & \makecell{\textbf{I/O}\\\textbf{Blocks}} \\
        \hline
        \multirow{3}{*}{\makecell{\textbf{Model}\\\textbf{Saving}}} & PostgresML & 4.828 & 2.491 & 1.276 & 0.780 & 1154816 \\
        & ELF$^\ast$ & 4.794 & 2.783 & 1.633 & 0.921 & 506776 \\
        & NeuralStore & 3.283 & 2.087 & 0.971 & 0.932 & 348504 \\
        \hline
        \multirow{3}{*}{\makecell{\textbf{Model}\\\textbf{Loading}}} & PostgresML & 1.919 & 0.503 & 1.308 & 0.944 & 702680 \\
        & ELF$^\ast$ & 1.982 & 1.395 & 0.583 & 0.998 & 507192 \\
        & NeuralStore & 0.895 & 0.470 & 0.227 & 0.779 & 348568 \\
        \hline
    \end{tabular}
    }
\end{table}

\subsection{Compression Performance Evaluation}

We now assess the compression performance of \dbname.
We conduct the experiments by progressively increasing the number of stored models from 200 to 800, and measure the storage consumption.
Moreover, we calculate the compression ratio according to the original size, totaling 361GB.

\subsubsection{Storage and Compression Ratio.}
We plot the storage sizes of the compressed models obtained using different compression algorithms in Figure~\ref{fig:eval_storage}.
\dbname consistently achieves the lowest compressed size across all model scales.
At 800 models, it reduces the total storage to 261GB, corresponding to a compression ratio of 1.38$\times$. While the compression ratios for ELF, ZFP, and ZSTD are 1.32$\times$, 1.18$\times$, and 1.10$\times$, respectively.
This trend persists across different scales, demonstrating the scalability and effectiveness of delta quantization compression.

\subsubsection{Per-model Compression Ratio Distribution.}

We study the per-model compression effectiveness of \dbname.
To account for shared base tensors, we evenly distribute the storage cost of each base tensor in the index across all tensors that reference it. 
Figure~\ref{fig:cdf_storage} shows the cumulative distribution function (CDF) of per-model compression ratios. 
\dbname outperforms all baselines across the distribution. Over 60\% of models achieve a compression ratio greater than 1.4$\times$, and nearly 90\% exceed 1.3$\times$. 
In contrast, no model compressed with ELF reaches
1.4$\times$, and fewer than 3\% of models do so with ZFP or ZSTD.
We also observe that the three baseline methods exhibit steep CDF curves concentrated between 1.2$\times$ and 1.3$\times$, which indicates limited variability in their compression effectiveness. For example, ELF compresses tensors by deduplicating the 8-bit exponent field of floating-point values, which caps its ideal compression ratio around 1.33$\times$. In contrast, \dbname exploits tensor-level similarities across models, enabling adaptive compression that achieves higher ratios when redundancy is present.

\extended{
\begin{figure}
\centering
\begin{minipage}{\linewidth}
    \centering
    \subfigure[Compression Ratio]{
        \label{fig:eval_tolerance_cr}
        \includegraphics[width=0.47\linewidth]{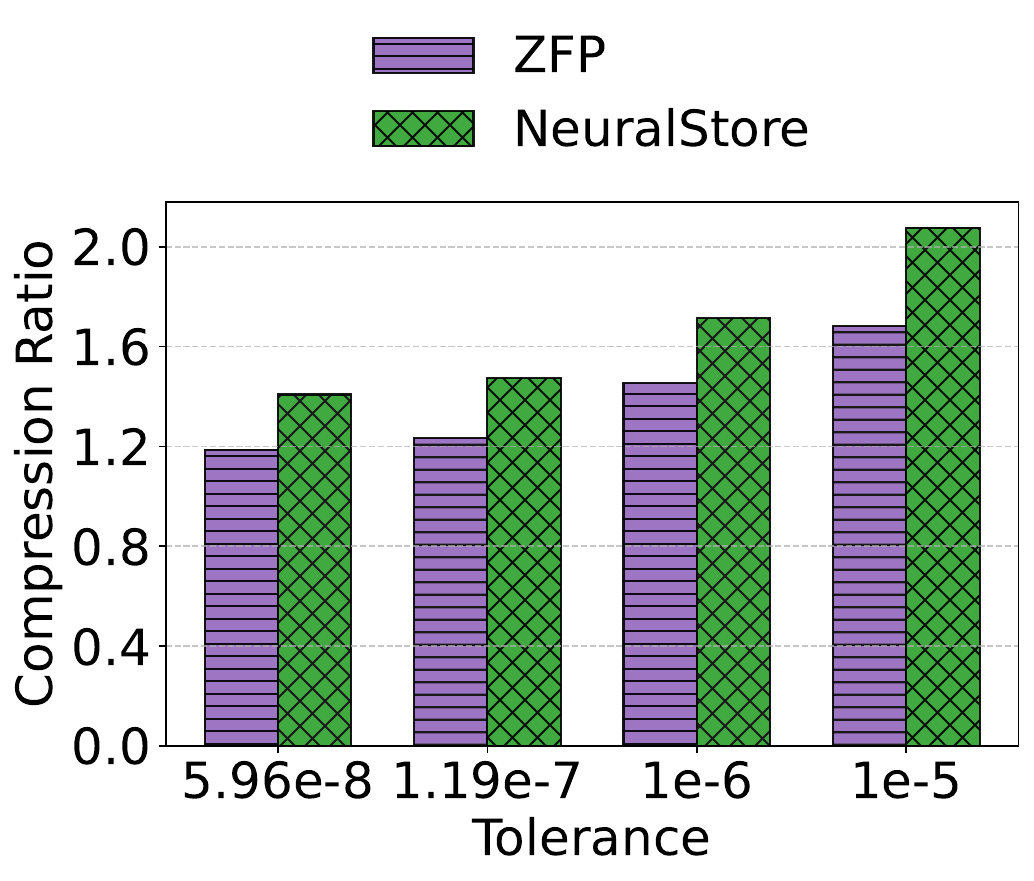}
    }
    \subfigure[Compression Throughput]{
        \label{fig:eval_tolerance_throughput}
        \includegraphics[width=0.47\linewidth]{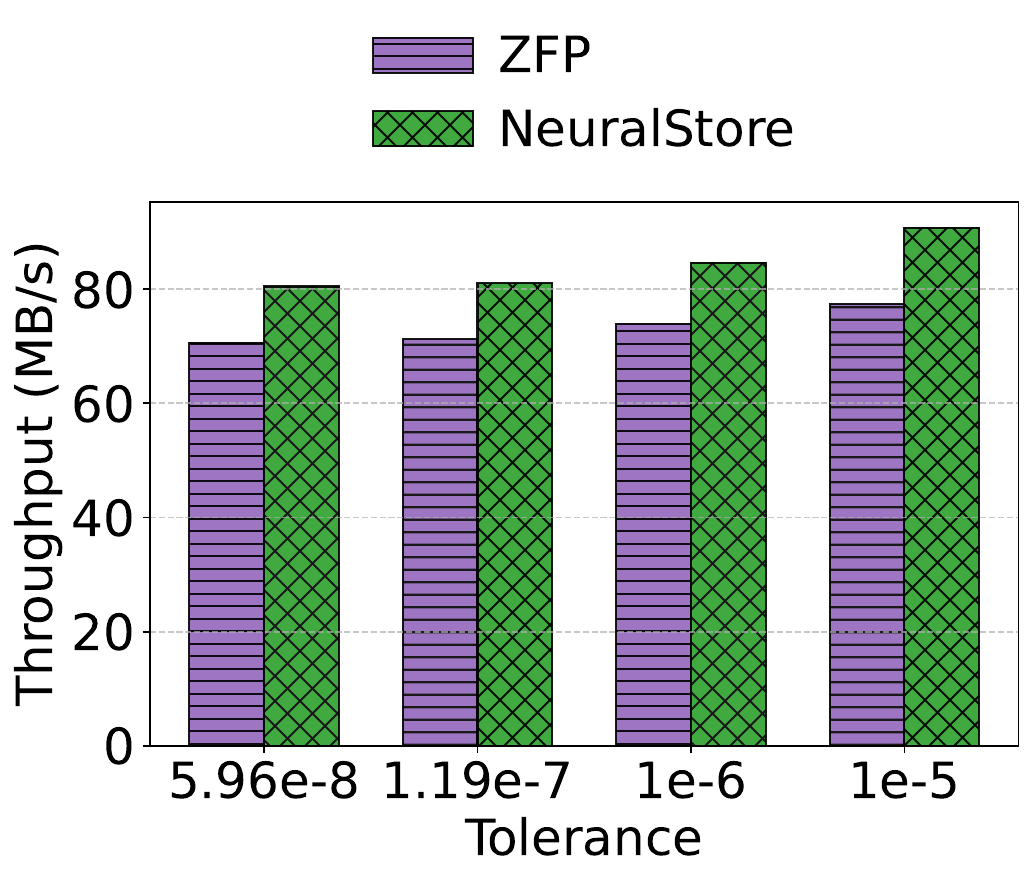}
    }
    \vspace{-4mm}
    \caption{Performance Impact of Precision Tolerance.}
    \vspace{-4mm}
    \label{fig:eval_tolerance}
\end{minipage}
\end{figure}
}

\begin{figure}
\begin{minipage}{\linewidth}
    \centering
    \subfigure[Model Loading Throughput]{
        \label{fig:eval_flexible_model_loading_throughput}
        \includegraphics[width=0.47\linewidth]{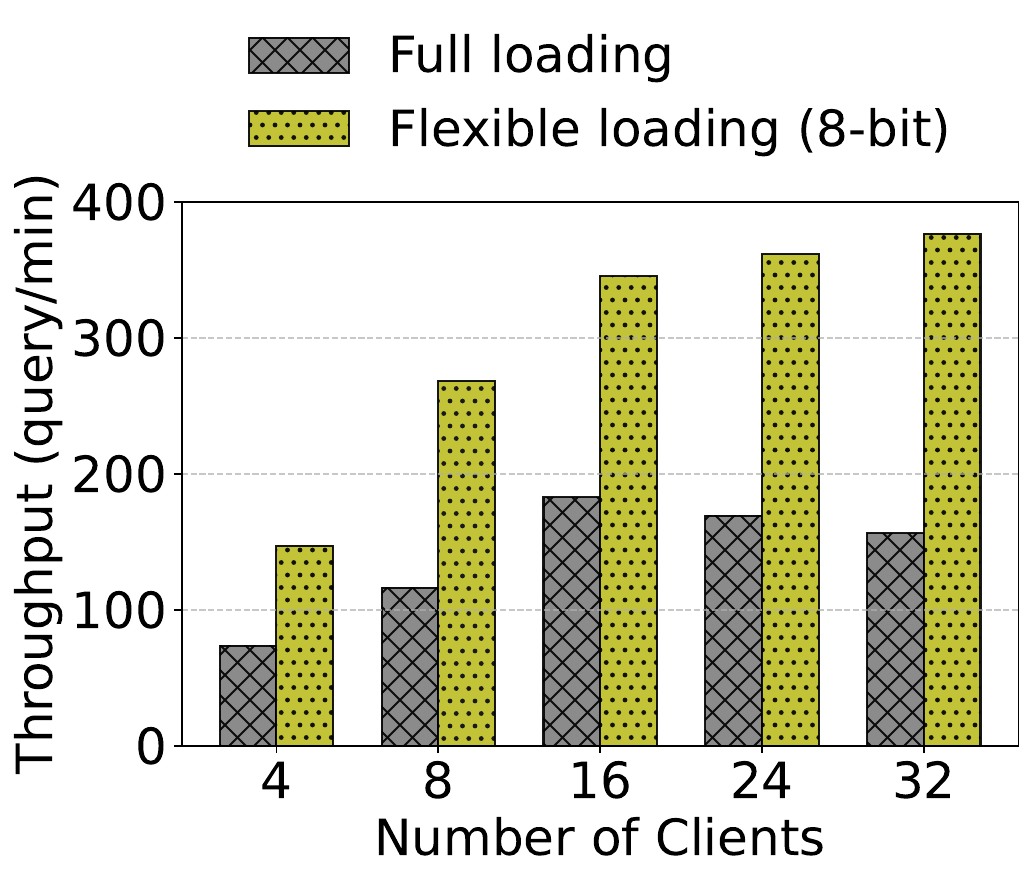}
    }
    \subfigure[Compression-aware Model Loading]{
        \label{fig:eval_flexible_model_loading_discard_bits}
        \includegraphics[width=0.47\linewidth]{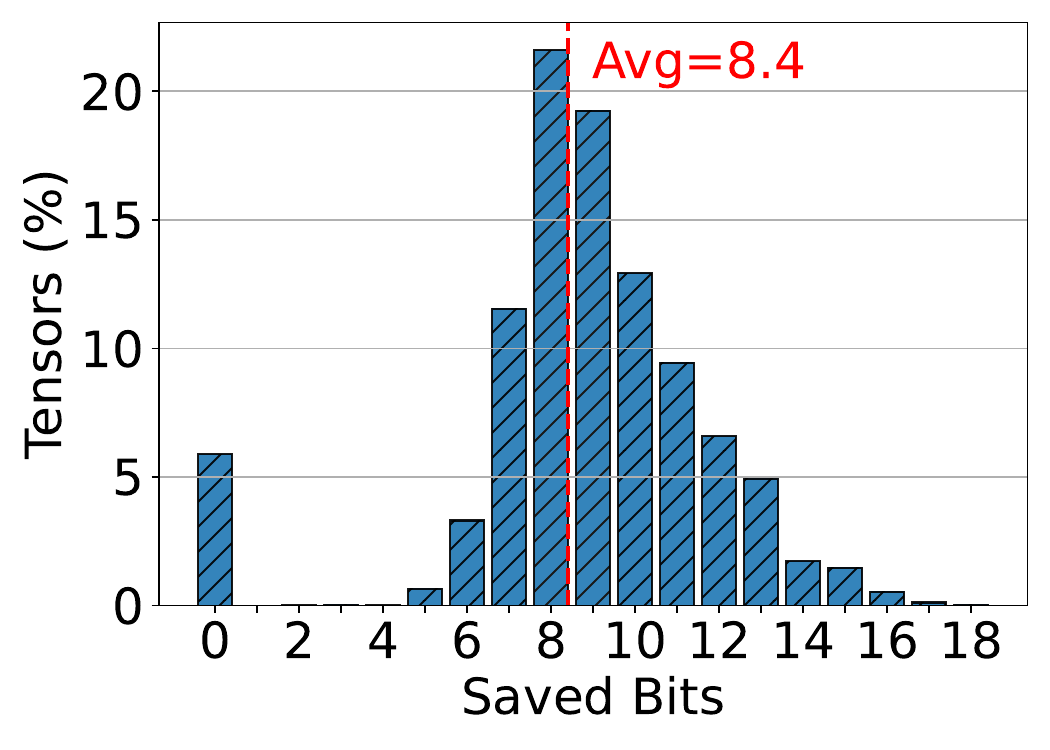}
        
    }
    \vspace{-4mm}
    \caption{Performance Impact of Flexible Model Loading.}
    \label{fig:eval_decompression_free}
\end{minipage}
\end{figure}

\subsection{Micro-Benchmarks}
\label{subsec:eval_micro}

To gain deeper insights into the performance trade-offs, we conduct micro-benchmarks to evaluate the impact of key system parameters, namely, the similarity threshold $\tau$, the user-defined precision tolerance $p$, as well as the design choice of flexible model loading.

\subsubsection{Performance Impact of Similarity Threshold}
\label{subsec:eval_sim_threshold}

A key parameter that influences the storage cost in \dbname is the similarity threshold ($\tau$), which decides whether a tensor can be delta-compressed depending on its distance from base tensors. 
A higher threshold enables more aggressive delta compression by accepting looser matches, thereby reducing the number of base tensors added to the HNSW indexes. However, this also results in degraded effectiveness of delta-encoding.
We evaluate the impact of $\tau$ using a subset of the dataset, consisting of 50 DL models fine-tuned from \texttt{google/bert-base}. We vary the similarity threshold and measure the resulting storage sizes of delta tensors and HNSW indexes.

Figure~\ref{fig:eval_tensor_based_storage_storage} shows how varying $\tau$ affects storage consumption. As $\tau$ increases, more tensors are qualified for delta encoding. Consequently, it reduces the opportunity of creating new vertex nodes in HNSW indexes, resulting in smaller index sizes.
However, the decrease in the storage of HNSW indexes slows down when $\tau$ increases beyond 0.16.
This is because HNSWs still need to maintain a minimum number of vertices for excessively distant tensors.
Similarly, the storage of delta tensors increases when $\tau$ is small, while the marginal increase diminishes as $\tau$ becomes higher.
It is because a higher similarity threshold results in a delta computed against sub-optimal base tensors, and therefore increases the number of bits required to represent the tensor.
When $\tau$ exceeds a certain range, in this case is 0.16, the allowed distance is greater than the nearest base tensors. As a result, tensors are always able to find another base tensor to create a smaller delta.
Therefore, the increase in the delta storage diminishes.
The overall compression ratio (plotted as a red line) reflects the trade-off between index storage and delta storage.
In the beginning, the index storage drops faster, leading to an increased compression ratio. It peaks at $\tau = 0.16$. 
After that, the increase in delta tensors storage overwhelms the space saved by indexes, leading to compression ratio drops.

We also evaluate the impact of $\tau$ on compression throughput under single-threaded and multi-threaded settings.
For the multi-threading setup, we perform compression on the same 50 BERT models using two threads.
Each thread fetches the tensor to be compressed from a shared queue and performs the similarity search, delta encoding, and quantization independently.
The results are shown in Figure~\ref{fig:eval_tensor_based_storage_throughput}.
As the similarity threshold $\tau$ becomes higher, the compression throughput increases from 90.5MB/s to 139.3MB/s for single-thread execution, and 155.8MB/s to 197.9MB/s for multi-thread execution.
This trend aligns with earlier observations, as more tensors are delta-compressed instead of inserted into the index, the system avoids costly HNSW index maintenance, leading to faster overall compression.
Notably, when $\tau = 0.16$, \dbname achieves the highest throughput and compression ratio. 


\extended{
\subsubsection{Performance Impact of Precision Tolerance}
\label{subsec:eval_quant}

We now evaluate the effectiveness of the delta quantization algorithm under varying precision tolerance $p$.
The precision tolerance is given by users as a parameter when storing each model.
It defines the upper bound of the quantization bin width, ensuring the resulting model accuracy is not significantly compromised.
Varying the precision tolerance enables users to balance the trade-off between compression performance and model accuracy.
In this experiment, we vary the precision tolerance from $5.96\times10^{-8}$ (single precision machine epsilon) to $10^{-5}$, and compare the compression ratio and throughput of \dbname and ZFP on the full 800-model set we collected.

The compression ratios are shown in Figure~\ref{fig:eval_tolerance_cr}.
As the precision tolerance increases, the compression ratio also improves because a wider tolerance (larger bin width) reduces the number of bits required during quantization.
\dbname consistently outperforms ZFP across all tolerance levels.
For example, at a tolerance of $10^{-5}$, the compression ratio of \dbname is 2.07$\times$ and 1.68$\times$ for ZFP, exhibiting a 1.2$\times$ improvement.
This is attributed to \revfour{\marginnote{R4.M3}\dbname's} ability to exploit inter-model tensor similarity, which becomes more effective as the precision tolerance increases.

We further measure the compression throughput with respect to the range of precision tolerance.
As illustrated in Figure~\ref{fig:eval_tolerance_throughput}, the throughput of \dbname increases from 80.4MB/s to 90.7MB/s as the tolerance varies from $5.96\times10^{-8}$ to $10^{-5}$.
This is because as precision tolerance increases, the number of bits to represent the delta becomes fewer.
Consequently, more tensors will fall within the similarity threshold, resulting in fewer tensors to be inserted into the HNSW indexes. 
By reducing the costly index insertions and graph maintenance operations, the overhead of delta quantization compression is significantly mitigated.
As a result, \dbname achieves faster compression rates at higher tolerance levels, without sacrificing its compression advantage.
}

\begin{figure}[t]
\centering
\includegraphics[width=0.99\linewidth]{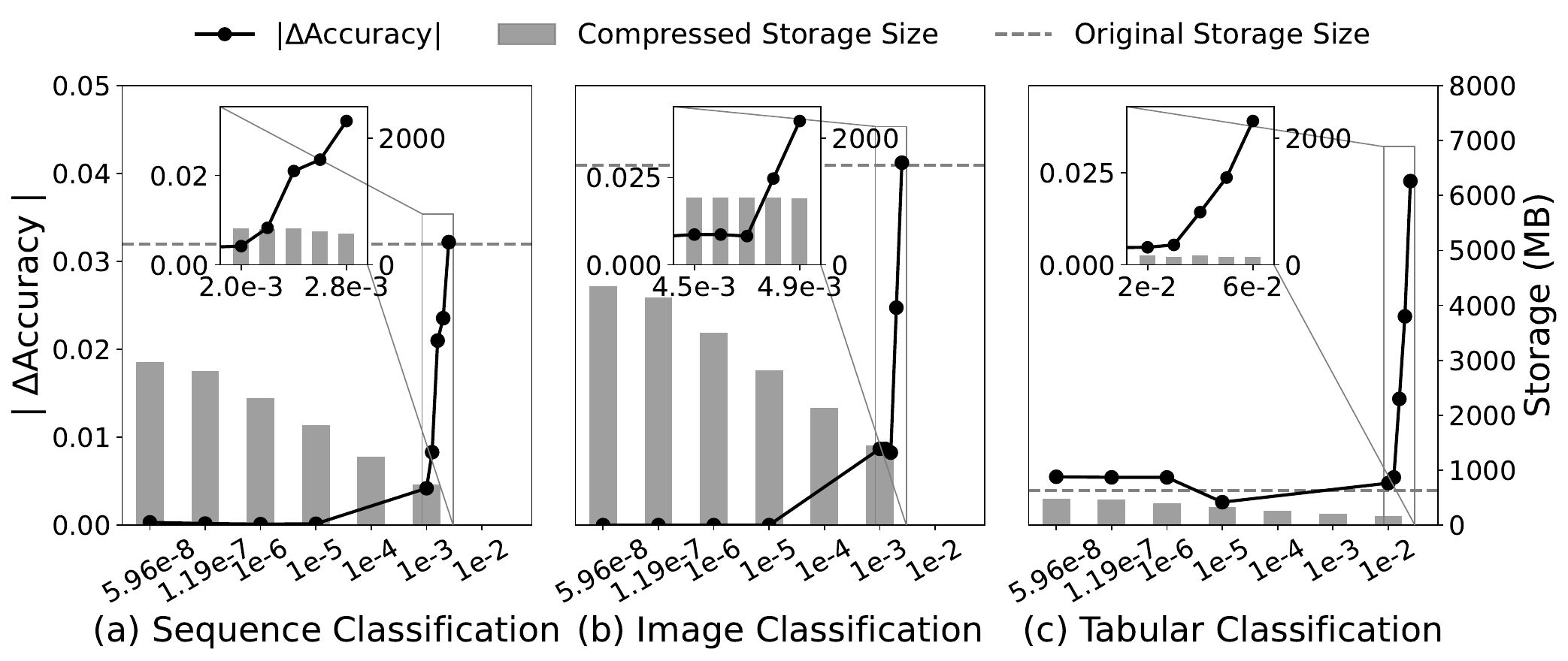}
\vspace{-4mm}
\caption{
Model Accuracy and Storage Change under Different Precision Tolerance
}
\label{fig.model_performance.accuracy_storage}
\end{figure}

\begin{figure*}[t]
\centering
\includegraphics[width=0.99\linewidth]{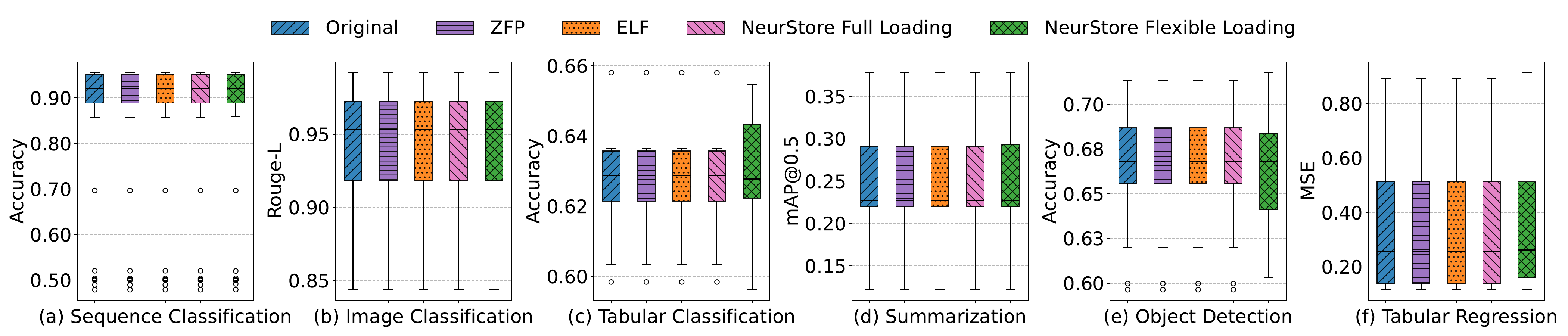}
\vspace{-4mm}
\caption{
Model Performance of Compared Compression Algorithms under Different Tasks
}
\vspace{-4mm}
\label{fig.model_performance.absolute}
\end{figure*}


\begin{figure}
\centering
\includegraphics[width=\linewidth]{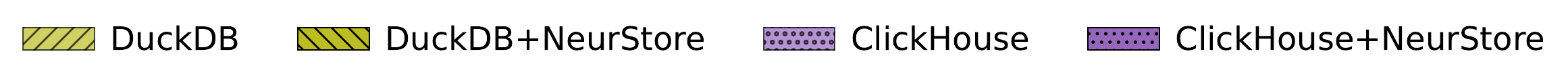}
\par\vspace{-2mm}
\subfigure[Write Throughput]{
\label{fig:duck_write_throughput}
\includegraphics[width=0.47\linewidth]{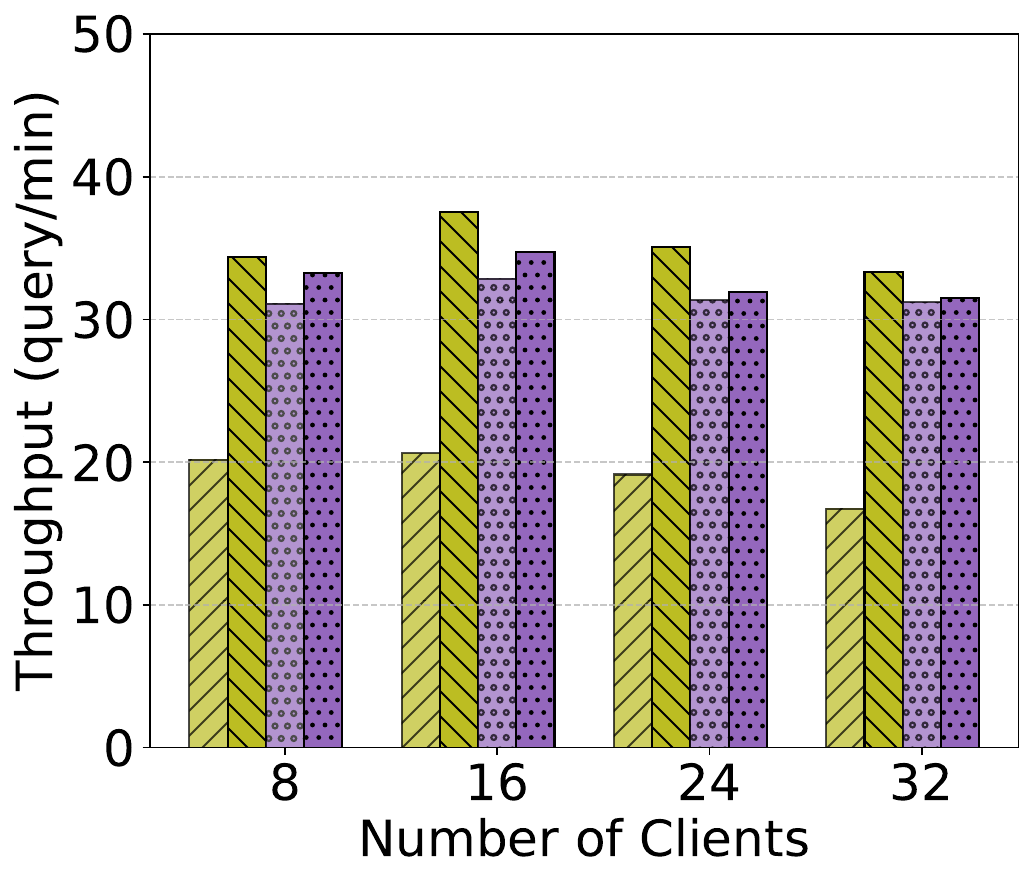}
}
\subfigure[Read Throughput]{
\label{fig:duck_read_throughput}
\includegraphics[width=0.47\linewidth]{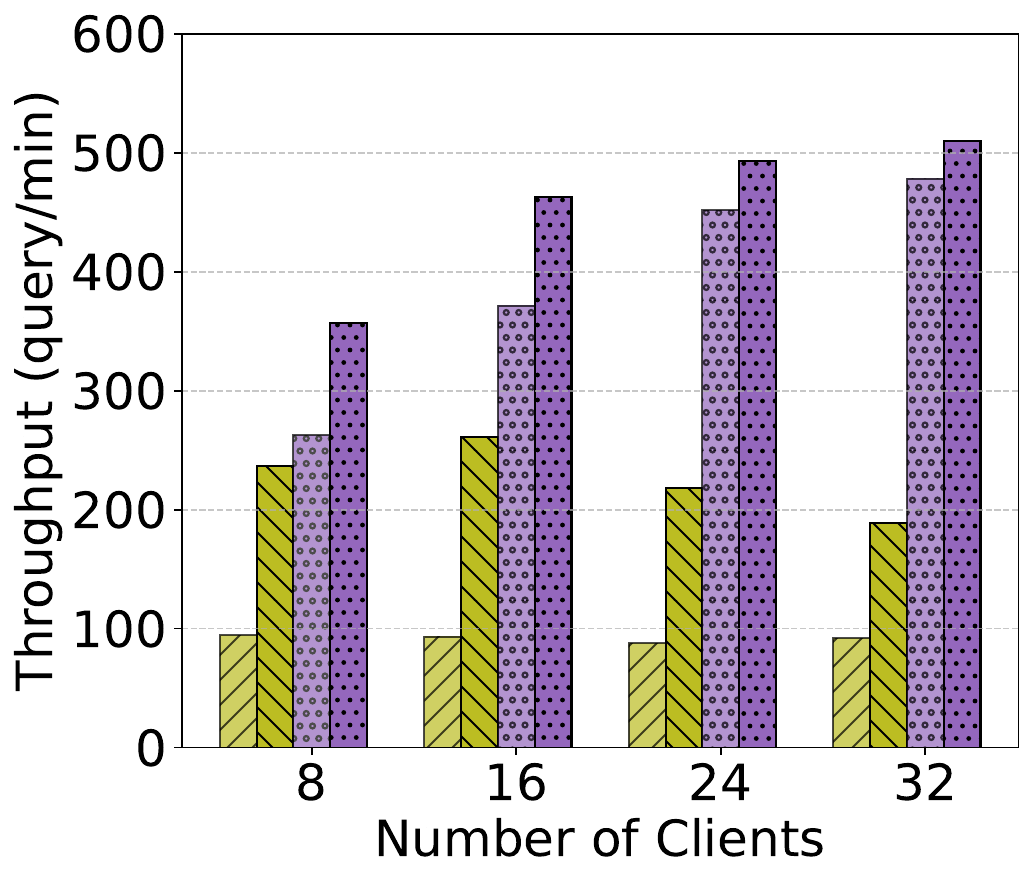}
}
\vspace{-4mm}
\caption{
Performance of \dbname on DuckDB and ClickHouse
}
\label{fig:eval_duckdb}
\end{figure}

\subsubsection{Performance Impact of Flexible Model Loading}
We evaluate the performance impact of flexible model loading, which provides users with options to load quantized-delta in full bit width and partial bit width.
To quantify the trade-off between model accuracy and efficiency in both model loading and memory usage, we conduct experiments with two loading strategies, namely full loading and flexible loading with 8 bits.

First, we measure the throughput of the two strategies.
The experiment is conducted by first initializing \dbname with 50 models using a precision tolerance of $5.96 \times 10^{-8}$.
We then run 4 to 32 clients, each of which continuously sends the load model queries.
Depending on the strategy, \dbname either loads full-bit-width or 8-bit delta tensors.
The results are presented in Figure~\ref{fig:eval_flexible_model_loading_throughput}.
It shows that 8-bit flexible loading achieves up to a 2.4$\times$ speedup compared to full loading.
This improvement is attributed to the reduced number of bits that need to be read, which lowers both disk I/O and memory usage.
Notably, the disk I/O becomes a bottleneck when models are concurrently retrieved by over 16 clients.
In contrast, the flexible loading strategy avoids such a bottleneck by significantly reducing the bits loaded.

Second, we evaluate the number of bits saved for each delta tensor using the flexible loading strategy.
To run the experiment, we initialize 800 models in \dbname, and load the models non-repeatedly.
We record the number of bits saved for each tensor and display the results in Figure~\ref{fig:eval_flexible_model_loading_discard_bits}.
It shows that flexible loading saves 8.4 bits on average, which indicates a compression ratio of 1.53$\times$.
With a precision tolerance of $5.96 \times 10^{-8}$, discarding this number of bits results in a precision loss of less than $10^{-4}$.
5\% of the delta tensors save 0 bits with flexible loading because their bit width is less than or equal to 8.
These tensors mainly originate from the same base tensor as their corresponding delta tensors.

\extended{
\begin{table}[t]
    \centering
    \revone{
    \caption{\revone{Summary of Models Used for Performance Change Evaluation}}
    \label{tab:performance_model}
    \vspace{-2mm}
    \resizebox{0.99\columnwidth}{!}{%
    \begin{tabular}{cccc}
        \hline
        \textbf{Domain} & \textbf{Task} & \textbf{Dataset} & \textbf{Architecture} \\
        \hline
        \multirow{6}{*}{NLP} 
        & \multirow{3}{*}{Sequence Classification} 
            & \multirow{3}{*}{IMDB} & BERT (9) \\
        &  &  & DistilBERT (22) \\
        &  &  & RoBERTa (16) \\
        \cline{2-4}
        & \multirow{3}{*}{Summarization} 
            & \multirow{3}{*}{SAMSum} & T5-small (42) \\
        &  &  & T5-base (20) \\
        &  &  & BART-base (10) \\
        \hline
        \multirow{5}{*}{CV} 
        & \multirow{4}{*}{Image Classification} 
            & Stanford Dogs & \multirow{4}{*}{\makecell{ViT-base (34)\\Swin-base (4)\\Swin-tiny (5)}} \\
        &  & Beans &  \\
        &  & Food-101 &  \\
        &  & CIFAR-10 &  \\
        \cline{2-4}
        & Object Detection & CPPE-5 & DETR-ResNet-50 (18) \\
        \hline
        \multirow{3}{*}{Tabular} 
        & Classification & Avazu & MLP (12) \\
        \cline{2-4}
        & \multirow{2}{*}{Regression} & \multirow{2}{*}{Regression} & MLP (4) \\
        &  &  & TabNet (4) \\
        \hline
    \end{tabular}
    }
    }
    \vspace{-4mm}
\end{table}
}

\subsubsection{Performance Impact of Precision Tolerance}
\label{subsec:eval_quant}
We then evaluate the impact of precision tolerance on model performance. We use the tasks and models as described in Section~\ref{subsec:workloads}. For each task, we gradually increase the precision tolerance from $5.96 \times 10^{-8}$ until a surge in models' average absolute performance change is observed. 
At each precision tolerance, we measure the average absolute accuracy change and compressed storage size. The results are shown in Figure~\ref{fig.model_performance.accuracy_storage}.
Across all tasks, increasing the precision tolerance leads to reduced storage consumption, as fewer bits are needed during quantization. However, the sensitivity to precision tolerance varies across tasks.
For sequence classification, model performance change remains within 0.02\% at tolerances below $1 \times 10^{-5}$. The model performance change begins to amplify from a tolerance of $2 \times 10^{-3}$, from where it increases from 0.42\% to 3.22\% at the tolerance of $2.8 \times 10^{-3}$.
For image classification, model performance remains unaffected at tolerances below $1 \times 10^{-5}$, but starts increasing from $4.5 \times 10^{-3}$, reaching a peak change of 4.12\% at $4.9 \times 10^{-3}$.
Lastly, tabular classification models maintain a performance change below 0.6\% until the tolerance reaches $2 \times 10^{-2}$, beyond which model performance change increases up to 3.91\% at $6 \times 10^{-2}$.
Users can configure a higher tolerance on a per-model basis to balance the trade-off between storage consumption and model performance degradation.
For example, models for image classification tasks can safely use a precision tolerance up to $1 \times 10^{-5}$, since no performance change is observed in Figure~\ref{fig.model_performance.accuracy_storage} at this tolerance level.

\subsubsection{Performance Impact Across Models}
\label{subsec:eval_model_performance}
We now evaluate the impact of flexible model loading and precision tolerance on model performance.
In addition to the three tasks introduced in Section~\ref{subsec:workloads}, we further include three more tasks to cover a wider range of model architectures: (1) summarization, where models generate abstractive summaries from dialogues, (2) object detection, where models identify multiple objects within images, and (3) tabular regression, where models predict continuous numeric values based on structured features.
In total, our evaluation covers 200 models across six tasks.
We compress and decompress these models using a fixed precision tolerance of $5.96\times10^{-8}$, and measure their absolute performance change. The results are shown in Figure~\ref{fig.model_performance.absolute}.
For ZFP, ELF, and \dbname Full Loading, over 90\% of models exhibit no performance change. For \dbname Flexible Loading, 57 out of 200 models show a performance change less than 0.01\%, and over 70\% of models exhibit a change within 0.1\%. 
More than 95\% of models remain within a performance change of 1\%. The increase in performance change is expected, since flexible loading restores only the most significant 8 bits of each delta tensor.
Among all tasks, object detection models are most sensitive to precision loss, with an average performance change of 0.8\%.
This is because the object detection task requires predicting bounding boxes in images, where small changes in model weights can lead to high deviations in predicted object locations.
On the contrary, image classification models show the smallest performance change, with an average of 0.009\%, due to the simplicity of the task.

\extended{
\begin{figure}
\centering
    \includegraphics[width=0.95\linewidth]{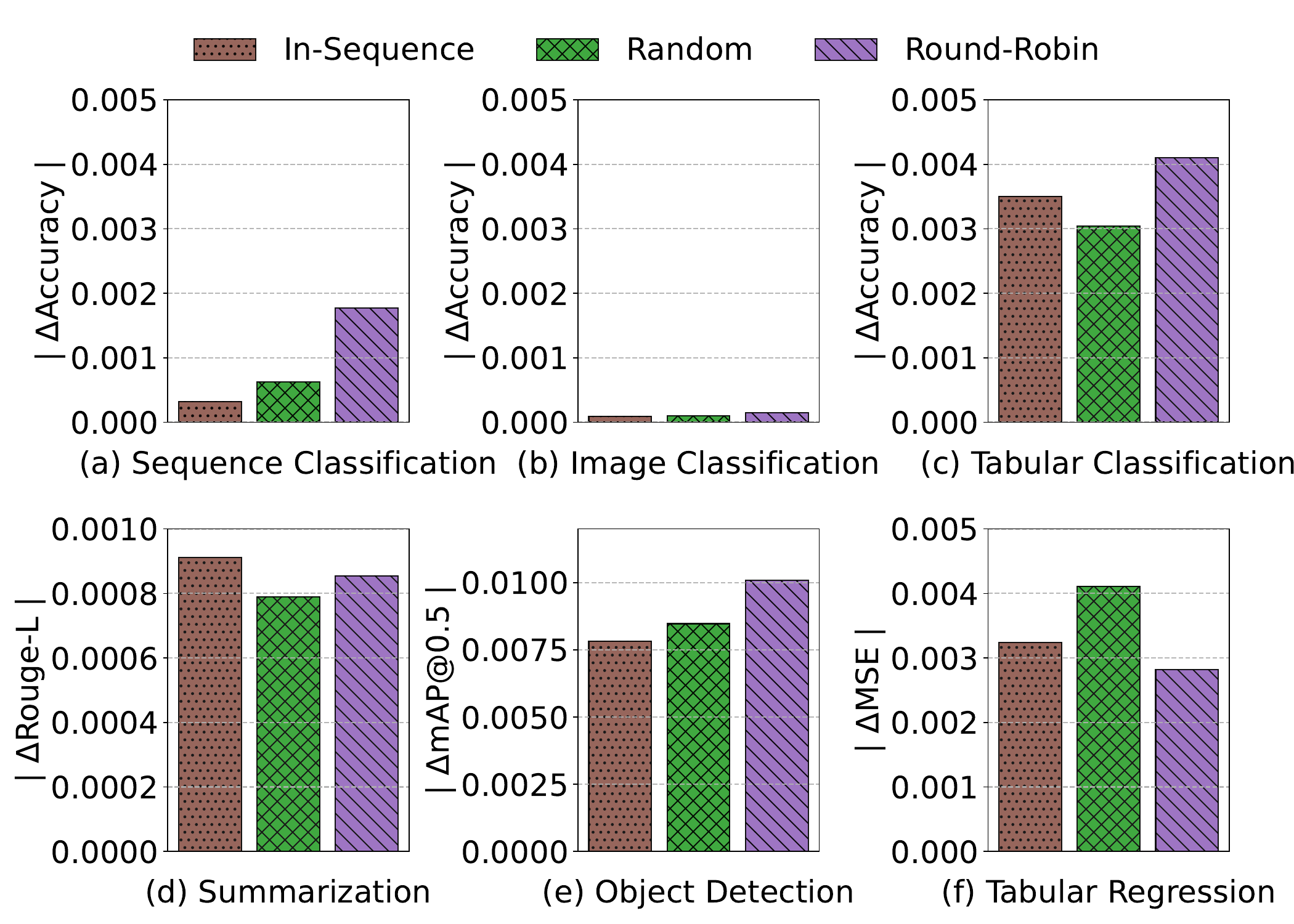}
\vspace{-4mm}
\caption{Model Performance Change under Different Orders}
\label{fig:eval_order}
\end{figure}

\subsubsection{Performance Impact of Storage Order}
To assess the impact of different storage orders on model accuracy, we insert models in three different orders and evaluate their resulting performance changes. Specifically, we evaluate the following storage orders:
1) Random order, which is the default setting we use to avoid potential bias from fixed storage sequences;
2) Sequential order, where models fine-tuned from the same pre-trained model are stored consecutively, representing an optimal case;
3) Round-robin order, which alternates storage across different architectures to maximize randomness and complexity, representing a worst-case scenario.
As shown in Figure~\ref{fig:eval_order}, the insertion order can affect model performance.
For the sequence classification, image classification, tabular classification, and object detection tasks, the round-robin order results in the highest model performance changes of 0.17\%, 0.01\%, 0.41\%, and 1.00\%, respectively.
For tabular regression, the random order causes the highest change, while for summarization, the sequential order yields the highest change.
Such performance deviations occur because tensors inserted earlier are more likely to be selected as bases in the similarity index.
As a result, changing the insertion order can lead to different base-delta pairings during compression and thus slightly influence model performance.
However, the overall effect of storing orders on model performance remains limited, as \dbname guarantees that the precision loss for each model stays within the user-defined tolerance.
}

\subsection{
Extensibility of \dbname
\label{subsec:eval_duckdb}}

We now extend \dbname to DuckDB~\cite{DBLP:conf/sigmod/RaasveldtM19}, denoted as DuckDB+\dbname, and to ClickHouse~\cite{DBLP:journals/pvldb/SchulzeSYDM24}, denoted as ClickHouse+\dbname.
For comparison, the baseline DuckDB and ClickHouse systems store each model by serializing it and applying Zstd compression before persisting on disk. All systems are configured with a 32 GB memory limit. For DuckDB, we use the persistent mode to fit our scenario with large model sizes. We vary the number of clients from 8 to 32 and measure both write and read throughput.
The results are shown in Figure~\ref{fig:eval_duckdb}. For write throughput, both DuckDB+\dbname and ClickHouse+\dbname consistently outperform their respective baselines across all numbers of clients. DuckDB+\dbname achieves a peak throughput of 38 queries per minute with 16 threads, which is 1.82$\times$ higher than DuckDB. This is due to the fact that \dbname reduces the I/O cost by utilizing the proposed tensor pages and HNSW index caching. ClickHouse+\dbname reaches a peak write throughput of 35 queries per minute at 16 threads, compared to 33 for ClickHouse. The higher baseline performance of ClickHouse compared to DuckDB can be attributed to its efficient data ingestion pipeline and background merge processes, which are inherently more efficient for large-block writes.
For read throughput, \dbname also outperforms the baselines across all levels of parallelism. DuckDB+\dbname achieves a peak throughput of 261 queries per minute, compared to 94 for DuckDB. These gains stem from \dbname’s in-memory index caching, on-demand decompression, and flexible model-loading strategy. ClickHouse+\dbname reaches a peak throughput of 510 queries per minute, compared to 478 for ClickHouse. The relative improvement of \dbname on ClickHouse is smaller because ClickHouse optimizes read throughput through uncompressed memory caching and block prefetching. In contrast, DuckDB relies on single-process execution with limited inter-query parallelism. As a result, \dbname yields larger performance improvements on DuckDB.
In addition, both DuckDB+\dbname and ClickHouse+\dbname consume only 78\% of the storage used by their respective baselines, demonstrating the overall effectiveness of our approach in reducing storage consumption.

\section{Related Works} \label{sec:related}


\paragraph{In-database Machine Learning}
Recently, there has been a growing interest in in-database machine learning (ML), which aims to integrate model training and inference directly within database engines to minimize data movement and exploit database-native execution for scalable analytics.
Early systems such as Bismarck~\cite{DBLP:conf/sigmod/FengKRR12}, MADlib~\cite{DBLP:journals/pvldb/HellersteinRSWFGNWFLK12}, and Oracle Machine Learning~\cite{oracle} embed learning algorithms into SQL-based workflows to enable large-scale model training over relational data.
More recent efforts, such as InferDB~\cite{DBLP:journals/pvldb/SalazarDiazGR24}, RAVEN~\cite{RAVEN_SIGMOD22}, CorgiPile~\cite{DBLP:conf/sigmod/XuQYJRGKLL0Y022}, Vertica-ML~\cite{VerticaML_SIGMOD2020}, push model inference into the database engine, optimizing the runtime serving path by tightly coupling inference with data access.
In addition, systems like EVA~\cite{EVA_SIGMOD22} and VIVA~\cite{VIVA_VLDB2022} enable declarative definition of machine learning pipelines for in-database video analytics.
NetsDB~\cite{netsDB_VLDB2022} proposes tensor deduplication during inference by identifying structural similarity across neural networks to improve inference efficiency.
While these systems primarily focus on in-database ML pipelines, they offer limited support for model management.
In contrast, \dbname enables efficient in-database model management, with a design tailored for modern DL models.
\extended{We introduce a set of techniques all natively embedded in the DBMS engine, including a tensor-based storage engine, adaptive delta quantization, and compression-aware model loading, to bridge the gap between model storage and inference within DBMSs.}





\paragraph{Model Management System}
MLCask~\cite{luo2021mlcask} is a model evolution management system that maintains the versioning of models in analytics pipelines, and it is complementary to \dbname.
Two earlier dedicated model management systems are
ModelDB~\cite{ModelDB_SIGMOD2016} and ModelHub~\cite{ModelHub_ICDE2017}.
ModelDB~\cite{ModelDB_SIGMOD2016} focuses on tracking model metadata and lineage to support machine learning workflow inspection, but it does not address model storage optimization.
To reduce storage overhead, ModelHub~\cite{ModelHub_ICDE2017} enables delta storage, which maintains the differences between fine-tuned and base models with explicit relations.
However, ModelHub only captures pairwise differences between models and their predecessors, and does not account for the high entropy of floating-point weights, which limits its storage efficiency.
In contrast, \dbname targets tensor-level deduplication across the entire model collection, while incorporating a delta quantization algorithm that can efficiently compress high-entropy delta tensors.
\section{Conclusion} \label{sec:conclusion}
This paper introduced \dbname, an efficient in-database deep learning model management system.
We introduced a tensor-based storage engine that enables fine-grained tensor deduplication by leveraging an enhanced HNSW-based tensor index.
To further reduce storage costs while preserving model performance, we proposed an adaptive delta quantization algorithm that dynamically compresses delta tensors with bounded accuracy loss.
Moreover, we designed a compression-aware loading and inference mechanism that supports direct computation on compressed tensors, significantly improving model retrieval and serving efficiency.
Extensive experimental results demonstrate that, compared to state-of-the-art in-database model management systems, \dbname achieves substantial storage savings while maintaining competitive model retrieval throughput and inference accuracy.



\bibliographystyle{ACM-Reference-Format}
\balance
\bibliography{main-bibliography}

\end{document}